\documentclass[journal,10pt]{IEEEtran}
\usepackage[T1]{fontenc}
\usepackage[utf8]{inputenc}
\usepackage{amsmath,amsthm,amssymb}
\usepackage{mathptmx,bm}
\usepackage{graphicx,booktabs,array,multirow,tabularx}
\usepackage[caption=false,font=footnotesize]{subfig}
\usepackage{float,stfloats}
\floatstyle{ruled}
\newfloat{algorithm}{tbp}{loa}
\floatname{algorithm}{Algorithm}
\makeatletter
\IfFileExists{algpseudocode.sty}{%
  \usepackage{algpseudocode}%
}{%
  \newcounter{ALG@line}%
  \newcounter{ALG@nest}%
  \newlength{\algorithmicindent}\setlength{\algorithmicindent}{1.05em}%
  \newcommand{\ALG@pad}{\hspace*{\numexpr\value{ALG@nest}\relax\algorithmicindent}}%
  \newenvironment{algorithmic}[1][0]{%
    \setcounter{ALG@line}{0}\setcounter{ALG@nest}{0}%
    \list{\footnotesize\arabic{ALG@line}:}{%
      \usecounter{ALG@line}%
      \setlength{\labelwidth}{1.35em}\setlength{\leftmargin}{1.65em}%
      \setlength{\labelsep}{0.3em}\setlength{\itemsep}{0.5pt}%
      \setlength{\parsep}{0pt}\setlength{\topsep}{1.5pt}}%
  }{\endlist}%
  \newcommand{\Require}{\item[\textbf{Require:}]}%
  \newcommand{\Ensure}{\item[\textbf{Ensure:}]}%
  \newcommand{\State}{\item\ALG@pad}%
  \newcommand{\If}[1]{\item\ALG@pad\textbf{if} ##1 \textbf{then}\stepcounter{ALG@nest}}%
  \newcommand{\ElsIf}[1]{\addtocounter{ALG@nest}{-1}\item\ALG@pad\textbf{else if} ##1 \textbf{then}\stepcounter{ALG@nest}}%
  \newcommand{\Else}{\addtocounter{ALG@nest}{-1}\item\ALG@pad\textbf{else}\stepcounter{ALG@nest}}%
  \newcommand{\EndIf}{\addtocounter{ALG@nest}{-1}\item\ALG@pad\textbf{end if}}%
  \newcommand{\For}[1]{\item\ALG@pad\textbf{for} ##1 \textbf{do}\stepcounter{ALG@nest}}%
  \newcommand{\EndFor}{\addtocounter{ALG@nest}{-1}\item\ALG@pad\textbf{end for}}%
  \newcommand{\Return}{\textbf{return} }%
}
\makeatother
\usepackage{cite,microtype,placeins,xspace,etoolbox,needspace}
\usepackage[nocheckfootnote]{flushend}
\usepackage[hidelinks]{hyperref}

\DeclareRobustCommand{\method}{CRISP\xspace}
\newcommand{\E}{\mathbb{E}}
\newcommand{\R}{\mathbb{R}}
\newcommand{\C}{\mathbb{C}}
\newcommand{\tr}{\operatorname{tr}}
\newcommand{\diag}{\operatorname{diag}}
\newcommand{\supp}{\operatorname{supp}}
\newcommand{\Cov}{\operatorname{Cov}}
\newcommand{\argmax}{\operatorname*{arg\,max}}
\newcommand{\argmin}{\operatorname*{arg\,min}}
\newcommand{\norm}[1]{\left\lVert#1\right\rVert}
\newcommand{\best}[1]{\textbf{#1}}
\newcommand{\second}[1]{\underline{#1}}

\newtheoremstyle{upright}{1.2pt}{1.2pt}{\normalfont}{}{\bfseries}{.}{.35em}{}
\theoremstyle{upright}
\newtheorem{theorem}{Theorem}

\newtheorem{lemma}{Lemma}
\newtheorem{corollary}{Corollary}

\let\itshape\upshape
\let\slshape\upshape
\AtBeginDocument{\let\itshape\upshape\let\slshape\upshape}
\renewcommand{\emph}[1]{#1}
\makeatletter
\renewcommand{\IEEEmembership}[1]{%
  {\@IEEEnotcompsoconly{\sublargesize}\normalfont
   \@IEEEcompsoconly{\sffamily}\fontshape{it}\selectfont #1}}
\makeatother

\newif\ifCRISPcompactlayout
\CRISPcompactlayouttrue
\ifCRISPcompactlayout
  \makeatletter
  \patchcmd{\section}{3.0ex plus 1.5ex minus 1.5ex}{0.9ex plus .12ex minus .12ex}{}{
    \PackageWarning{CRISP}{IEEEtran section-before skip was not patched}}
  \patchcmd{\section}{0.7ex plus 1ex minus 0ex}{.22ex plus .06ex minus .04ex}{}{
    \PackageWarning{CRISP}{IEEEtran section-after skip was not patched}}
  \patchcmd{\subsection}{3.5ex plus 1.5ex minus 1.5ex}{.85ex plus .10ex minus .10ex}{}{}
  \patchcmd{\subsection}{0.7ex plus .5ex minus 0ex}{.20ex plus .06ex minus .04ex}{}{}
  \def\@IEEENORMtitlevspace{1.1\baselineskip}
  \def\@IEEEMINtitlevspace{0.9\baselineskip}
  \expandafter\patchcmd\csname\string\proof\endcsname
    {\topsep6\p@\@plus6\p@}{\topsep1.2\p@\@plus.25\p@}{}{
    \PackageWarning{CRISP}{Proof spacing was not patched}}
  \patchcmd{\@makecaption}{\addvspace{0.5\baselineskip}}{\addvspace{0.22\baselineskip}}{}{}
  \makeatother
  \newcommand{\CRISPDisplaySkips}{%
    \setlength{\abovedisplayskip}{1.2pt plus .25pt minus .30pt}%
    \setlength{\belowdisplayskip}{1.2pt plus .25pt minus .30pt}%
    \setlength{\abovedisplayshortskip}{.4pt plus .2pt minus .2pt}%
    \setlength{\belowdisplayshortskip}{1.2pt plus .2pt minus .2pt}}
  \AtBeginDocument{%
    \apptocmd{\normalsize}{\CRISPDisplaySkips}{}{}%
    \CRISPDisplaySkips
    \setlength{\abovecaptionskip}{1.8pt}%
    \setlength{\belowcaptionskip}{0pt}%
    \setlength{\jot}{1pt}%
    \setlength{\IEEEnormaljot}{1pt}}
\fi

\newif\ifCRISPallowlooseness
\CRISPallowloosenesstrue

\ifCRISPcompactlayout
  \AddToHook{para/begin}{\ifCRISPallowlooseness\looseness=-1\relax\fi}
  \AtBeginEnvironment{tabularx}{\CRISPallowloosenessfalse}
  \AtBeginEnvironment{tabular*}{\CRISPallowloosenessfalse}
  \renewcommand{\IEEEbibitemsep}{.05pt plus .15pt}
  \AtBeginEnvironment{proposition}{\Needspace*{2\baselineskip}}
\fi

\newlength{\CRISProwheight}
\newcommand{\CRISPsetrowheight}[1]{%
  \settoheight{\CRISProwheight}{%
    \includegraphics[width=.318\textwidth]{#1}}}

\AtBeginEnvironment{theorem}{\Needspace*{2\baselineskip}}
\AtBeginEnvironment{lemma}{\Needspace*{2\baselineskip}}
\AtBeginEnvironment{corollary}{\Needspace*{2\baselineskip}}

\newcommand{\MatchedFullKRMSE}{50.02}
\newcommand{\MatchedSingleRMSE}{51.16}
\newcommand{\MatchedFusionGain}{2.23}

\newcommand{\MatchedTrialCount}{5\,376}
\newcommand{\CRISPMeanRuntime}{82.63}
\newcommand{\SnapshotMeanRuntime}{44.01}
\newcommand{\FitTrialCount}{1\,792}
\newcommand{\GuardTrialCount}{896}

\newcommand{\CoreTrials}{5\,376}
\newcommand{\CoreCRISPRMSE}{50.03}
\newcommand{\CoreSnapshotRMSE}{58.07}
\newcommand{\CoreRMSEGain}{13.84}

\newcommand{\CoreCRISPTailRate}{3.96}
\newcommand{\CoreSnapshotTailRate}{6.75}
\newcommand{\CoreTailReduction}{41.32}
\newcommand{\CoreGainCILow}{11.97}
\newcommand{\CoreGainCIHigh}{15.57}
\newcommand{\CoreMeanSolves}{1.43}
\newcommand{\FixedTrials}{2\,688}
\newcommand{\FixedCRISPRMSE}{50.40}
\newcommand{\FixedSnapshotRMSE}{58.01}
\newcommand{\FixedRMSEGain}{13.12}

\newcommand{\DisplacedTrials}{2\,688}
\newcommand{\DisplacedCRISPRMSE}{49.67}
\newcommand{\DisplacedSnapshotRMSE}{58.13}
\newcommand{\DisplacedRMSEGain}{14.57}

\newcommand{\GeometryTrials}{3\,584}
\newcommand{\GeometryCRISPRMSE}{61.22}
\newcommand{\GeometrySnapshotRMSE}{69.75}
\newcommand{\GeometryRMSEGain}{12.23}

\newcommand{\GeometryCRISPTailRate}{9.04}
\newcommand{\GeometrySnapshotTailRate}{12.00}
\newcommand{\GeometryTailReduction}{24.65}
\newcommand{\GeometryGainCILow}{10.08}
\newcommand{\GeometryGainCIHigh}{14.35}

\newcommand{\HoldoutTrials}{896}
\newcommand{\HoldoutCRISPRMSE}{45.06}
\newcommand{\HoldoutSnapshotRMSE}{49.68}
\newcommand{\HoldoutRMSEGain}{9.31}

\newcommand{\HoldoutTailReduction}{34.29}
\newcommand{\HoldoutGainCILow}{3.51}
\newcommand{\HoldoutGainCIHigh}{15.08}

\newcommand{\HoppingTrials}{768}
\newcommand{\HoppingCRISPRMSE}{203.58}
\newcommand{\HoppingSnapshotRMSE}{192.10}

\newcommand{\PaperObservationCount}{17\,568}
\newcommand{\PaperConditionCount}{201}

\newcommand{\RerouteCount}{3\,106}
\newcommand{\RerouteOffRMSE}{152.80}
\newcommand{\RerouteOnRMSE}{119.37}
\newcommand{\RerouteReduction}{33.43}
\newcommand{\BinTwoCoreRMSE}{82.80}
\newcommand{\BinTwoSingleRMSE}{87.47}
\newcommand{\BinTwoGain}{5.34}
\newcommand{\BinTwoGainCILow}{3.00}
\newcommand{\BinTwoGainCIHigh}{7.61}
\newcommand{\BinTwoTrials}{768}

\newcommand{\OverallMeanSolves}{1.51}
\newcommand{\RepFitTrialCount}{896}
\newcommand{\CoreCRISPClockRMSE}{30.53}
\newcommand{\CoreSnapshotClockRMSE}{35.78}

\hypersetup{
  pdftitle={CRISP: Sparse Risk-Calibrated Front-End Allocation and Fusion for Low-Resolution Snapshot GNSS Positioning},
  pdfauthor={Zhihan Zeng; Kaihe Wang; Zhongpei Zhang; Li-Ta Hsu}
}

\begin{document}
\title{CRISP: Sparse Risk-Calibrated Front-End\\ Allocation and Fusion for Low-Resolution\\ Snapshot GNSS Positioning}
\author{Zhihan Zeng, \IEEEmembership{Graduate Student Member, IEEE},
Kaihe Wang, \IEEEmembership{Graduate Student Member, IEEE},\\
Zhongpei Zhang, Li-Ta Hsu, \IEEEmembership{Senior Member, IEEE}
\thanks{Zhihan Zeng is with the National Key Laboratory of Wireless Communications, University of Electronic Science and Technology of China, Chengdu, China, e-mail: zzhihan@std.uestc.edu.cn.}%
\thanks{Kaihe Wang and Zhongpei Zhang are with the Shenzhen Institute for Advanced Study, University of Electronic Science and Technology of China, Shenzhen 518110, China, e-mail: khewang@yeah.net; zhangzp@uestc.edu.cn.}%
\thanks{Li-Ta Hsu is with The Hong Kong Polytechnic University, Hong Kong SAR, China, e-mail: lt.hsu@polyu.edu.hk.}
\thanks{Corresponding author: Zhongpei Zhang.}}
\maketitle

\begin{abstract}
Low-resolution global navigation satellite system snapshot receivers face strong coupling among automatic gain control compression, coarse quantization, and waveform distortion from nonlinear interference mitigation. This paper presents Cross-Risk Interference-Resilient Snapshot Positioning, denoted \method{}, an aided single-snapshot architecture that allocates and fuses complementary pre-correlation processors over a bounded position-clock search region under a shared direct position estimation solve budget. On a pre-frozen baseband descriptor partition, offline calibration estimates uncentered second-order position-error cross moments and solves cardinality-constrained simplex programs along east, north, and up axes, retaining validation-screened multibranch combinations while assigning single matched processors elsewhere. An online reliability gate screens the same quantized snapshot to route unfamiliar waveforms toward single-branch protective operators. We characterize profiled contrast shifts after nonlinear suppression and derive a conditional finite-sample excess-risk bound across disjoint splits alongside an exact same-snapshot gating risk identity. Across $\PaperObservationCount{}$ simulated $2$-ms GPS L1 C/A trials, \method{} reduces core positioning root-mean-square error by \CoreRMSEGain{}\%, $>100$-m gross errors by \CoreTailReduction{}\%, and multi-geometry error by \GeometryRMSEGain{}\% over Snapshot-FD at \CoreMeanSolves{} core and \OverallMeanSolves{} overall solves per snapshot.
\end{abstract}


\section{Introduction}\label{sec:introduction}
\IEEEPARstart{A}{irborne}, unmanned aerial, and power-constrained navigation platforms increasingly employ millisecond-scale baseband snapshots to obtain rapid position fixes during duty-cycled operation or fast reacquisition after jamming-induced loss of lock. Operating over only a few milliseconds of signal capture without tracking-loop convergence, short-snapshot receivers contend with limited spread-spectrum integration gain alongside coarse analog-to-digital conversion and block root-mean-square automatic gain control, denoted AGC. When strong interference enters the analog front end, it alters AGC scaling and reshapes quantizer level occupancy before digital filtering begins, compressing the dynamic range available to the desired satellite waveforms \cite{snapshot}.

Direct position estimation, denoted DPE, improves weak-signal resilience by fitting multi-satellite spreading replicas directly to a shared navigation and clock state \cite{dpe2007,dpecrb,Zeng2026PhysicsGuidedMoE,Yan2025PrincipalGaussianOverbound,Vicenzo2024GNSSCorrelogram,Vicenzo2023ExperimentalDPE}. Preceding DPE with nonlinear interference suppression introduces a fundamental trade-off between jammer rejection and waveform fidelity. Pre-correlation operators distort the desired mean waveform and modify the projection covariance of residual noise onto the satellite replica subspace. Time-domain, frequency-domain, cascaded, time-frequency, and spectral-line front ends offer complementary suppression mechanisms across continuous-wave, pulsed, and swept-frequency jammers \cite{Fang2025MultipathDopplerError,Fang2026MultipathDopplerEstimation,rim,snapshot,ridge,plse,Zeng2026DualStreamGNSSJamming}, so no single fixed processor dominates across diverse waveform regimes. Executing an entire bank of front ends in parallel multiplies the cost of multistart DPE optimization, while naive output averaging ignores strong error correlations and biases induced by sharing the same quantized snapshot.

Addressing this coupling requires formulating front-end selection, coordinate-level fusion, and computational budgeting as a unified statistical decision problem. Combining multiple predictors via simplex-constrained residual cross-moment quadratic programs originates in stacked regressions \cite{breiman1996stacked}; although zero-weight candidates need not be evaluated online, unpruned simplex fits without cardinality caps can retain multiple small nonzero weights and invoke many candidate branches per snapshot. This paper develops \method{}, which extends simplex residual-moment stacking to GNSS waveform-conditioned sparse front-end allocation under an explicit shared DPE solve budget. Within each pre-frozen waveform-descriptor bin, \method{} calibrates uncentered second-order position-error cross moments offline from paired known-position realizations, aligning the objective algebraically with uncorrected mean-squared position error, and solves convex simplex quadratic programs over candidate supports of at most four branches. An independent validation screen prunes multibranch combinations back to the bin-wise single-branch anchor whenever paired error reduction does not clear a standard-error margin, expending extra DPE solves only where residual errors compensate. An ordered online reliability gate operating on the same quantized observation screens descriptor support, low-lag correlation, and quantizer occupancy against frozen applicability thresholds to route unfamiliar waveforms toward single-branch protective processors.

The contributions are as follows:
\begin{enumerate}
\item A waveform-conditioned sparse front-end allocation and coordinate-simplex fusion architecture driven by calibrated three-dimensional position risk, extending simplex residual-moment stacking \cite{breiman1996stacked} to a shared DPE branch-execution budget on a single low-resolution quantized snapshot.
\item A theoretical characterization of profiled DPE contrast shift after nonlinear interference processing, paired with a conditional finite-sample excess-risk bound across disjoint representation-fit and risk-fit splits and an exact risk-accounting identity for same-snapshot gating.
\item A simulation evaluation over $\PaperObservationCount{}$ aided $2$-ms GPS L1 C/A trials spanning compound AGC-quantizer interference, changed satellite geometries, translated position centers, aiding-error sweeps, quantizer bit-depth sweeps, and fast frequency hopping, quantifying multibranch synergy, scalar versus coordinate weighting, accuracy-latency frontiers, and online gate operating boundaries.
\end{enumerate}

Section~\ref{sec:related} reviews DPE under waveform distortion, single-channel interference mitigation, and correlated-estimate fusion. Section~\ref{sec:system} defines the aided quantized snapshot model and shared profiled DPE backend. Section~\ref{sec:method} details the processor library, sparse cross-risk calibration, validation screen, and online reliability gate. Section~\ref{sec:theory} analyzes nonlinear contrast mismatch and finite-sample calibration bounds. Sections~\ref{sec:protocol} and~\ref{sec:results} present the experimental protocol and numerical analyses, and Section~\ref{sec:conclusion} concludes the paper.

\section{Related Work}\label{sec:related}
Direct position estimation embeds navigation coordinates directly into baseband correlation fitting \cite{dpe2007}, with Cram\'er-Rao analyses establishing its information advantage over two-step ranging under low post-correlation signal-to-noise ratios \cite{dpecrb}. Subsequent work has extended DPE to dense multipath channels \cite{dense}, multireceiver aerial platforms \cite{aerial}, navigation-domain geometric conditioning \cite{geometry}, grid-reduced data association \cite{robustdpe}, and software-defined receivers \cite{precise}. Whereas spatial anti-jamming techniques such as power-based Capon beamforming exploit multi-antenna degrees of freedom \cite{pbc} and parametric or neural correlators target multipath delay structure \cite{multipath,deepcorr}, single-antenna snapshot receivers must suppress high-power interference directly within a brief quantized scalar stream before evaluating a shared DPE objective.

Single-channel robust interference mitigation, denoted RIM, applies zero-memory or block nonlinearities across time, frequency, or cascaded domains \cite{rim}, with recent snapshot studies demonstrating strong sensitivity to scale estimation and subblock length under coarse quantization \cite{snapshot}. Under nonstationary frequency-swept jammers, time-frequency ridge tracking and piecewise least-squares phase modeling in the fractional Fourier transform, or FrFT, domain isolate concentrated chirp energy \cite{ridge,plse}, while compound spectral and time-frequency descriptors support waveform classification \cite{compound}. Rather than mapping descriptors to hard jammer-class labels that trigger one fixed filter, \method{} uses standardized baseband descriptors to condition a continuous position-risk model across complementary operators feeding a common DPE backend.

Combining estimators derived from a common observation raises the classical problem of correlated estimation errors \cite{barshalom}. In supervised learning, Breiman's stacked regressions \cite{breiman1996stacked} model cross-estimator dependence via nonnegative unit-simplex quadratic programs: whereas the population simplex oracle achieves risk no worse than the best individual predictor, empirical weight estimation incurs finite-sample variance that motivates support sparsity and out-of-sample validation. When cross-correlations are unknown online, covariance intersection and split-covariance rules bound worst-case dispersion \cite{ci,cros}, whereas shrinkage estimators regularize sample covariances \cite{shrink}. In nonlinearly preprocessed low-bit GNSS receivers, widely linear I/Q filtering captures improper complex interference \cite{picinbono,improper,widely}, while AGC coupling, coarse quantization, and adaptive thresholding induce downstream likelihood misspecification \cite{mismatch,scatter,delaymismatch}. Building on simplex stacking \cite{breiman1996stacked}, \method{} conditions cardinality-constrained ENU weights on low-bit baseband descriptors using offline known-position snapshots and couples shared-budget DPE execution with same-snapshot applicability gating.

\section{Observation Model and Receiver Objective}\label{sec:system}
We consider aided single-snapshot positioning \cite{vandiggelen2009agps} in a single-channel GPS L1 C/A receiver operating over a bounded four-dimensional position-clock search region $\Theta$ with known ADC bit depth $B$. Broadcast or network-assisted ephemerides and atmospheric models supply satellite positions $\boldsymbol p_{s,k}$ and range corrections $d_k$ to meter-level precision, while cellular, inertial, or pre-jamming navigation assistance provides a coarse reference state $\boldsymbol\theta_{\mathrm{ref}}$ that bounds true three-dimensional east-north-up coordinates and receiver clock bias within a $\pm 300$-m cube equivalent to $\pm 1\,\mu\text{s}$ in clock offset, defining an assisted acquisition regime rather than an unaided cold-start global search. Aided line-of-sight carrier Doppler and oscillator drift estimates maintain residual frequency offsets within the $500$-Hz coherent mainlobe of the $2$-ms capture spanning two consecutive $1$-ms C/A code periods. Online inference processes a single $N$-sample complex-baseband snapshot $\boldsymbol y$ to estimate the three-dimensional ENU position alongside the receiver clock bias, whereas ground-truth states serve exclusively during offline calibration and test evaluation.

\begin{figure*}[!t]
\centering
\includegraphics[width=.94\textwidth]{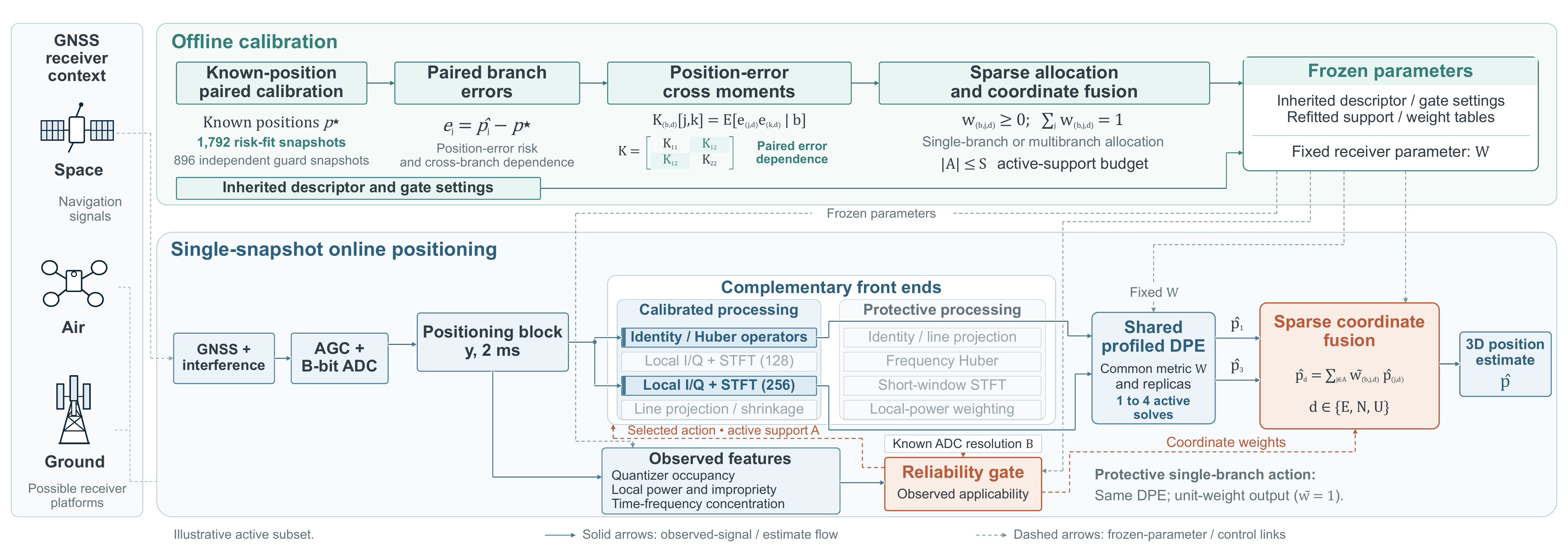}
\caption{Single-snapshot CRISP architecture for the $2$-ms GPS L1 C/A receiver configuration. Offline known-position data calibrate sparse coordinate-fusion policies across pre-frozen descriptor bins, while online feature extraction, reliability gating, active front ends, and shared profiled DPE solves operate on one quantized snapshot.}
\label{fig:architecture}
\end{figure*}

Let $\boldsymbol\theta=[\boldsymbol p^{\mathsf T},b_c]^{\mathsf T}\in\mathbb R^4$ collect the ENU position $\boldsymbol p$ and receiver clock bias $b_c$ in meters. Over one snapshot of $N$ complex baseband samples, the unquantized waveform under narrowband, pulsed, mixed, or swept-frequency interference $\boldsymbol j$ and circular complex Gaussian thermal noise $\boldsymbol n$ is
\begin{equation}\label{eq:waveform}
 \boldsymbol x=\boldsymbol C(\boldsymbol\theta_\star)\boldsymbol\alpha_\star
                 +\boldsymbol j+\boldsymbol n,
 \qquad\boldsymbol n\sim\mathcal{CN}(\boldsymbol0,\boldsymbol I_N).
\end{equation}
The $K$ columns of $\boldsymbol C(\boldsymbol\theta_\star)$ contain bandlimited spreading-code replicas modulated by aided Doppler carriers, while $\boldsymbol\alpha_\star\in\mathbb C^K$ collects unknown per-satellite complex channel gains and carrier phases. For satellite $k$ at position $\boldsymbol p_{s,k}$, propagation delay obeys
\begin{equation}\label{eq:delay}
 \tau_k(\boldsymbol\theta)=
 \frac{\|\boldsymbol p_{s,k}-\boldsymbol p\|+b_c+d_k}{c},
\end{equation}
with speed of light $c$. Code replicas are synthesized via exact geometric range differences and periodic Fourier interpolation over $\Theta$. Prior to quantization, block-RMS AGC normalizes $\boldsymbol x$ by $g(\boldsymbol x)=(N^{-1}\|\boldsymbol x\|^2)^{-1/2}$ to produce
\begin{equation}\label{eq:adc}
 \boldsymbol y=Q_B\{g(\boldsymbol x)\boldsymbol x\}.
\end{equation}
For $B>1$, real and imaginary parts use independent uniform $2^B$-level mid-rise quantizers over $[-3/\sqrt2,3/\sqrt2]$, whereas $B=1$ retains component signs. Signal quality and jammer intensity are parameterized by $C/N_0$ and pre-AGC snapshot-averaged J/N. Because desired signals, interference, and thermal noise share $g(\boldsymbol x)$, strong jamming compresses desired-signal amplitude and reshapes quantizer occupancy prior to filtering, as depicted in Fig.~\ref{fig:architecture}.

\subsection{A Common Profiled Positioning Backend}
Each candidate front-end processor $\mathcal R_j$ maps $\boldsymbol y$ to a suppressed length-$N$ waveform $\boldsymbol z_j=\mathcal R_j(\boldsymbol y)$ feeding a shared profiled DPE backend with positive definite spectral weighting
\begin{equation}\label{eq:metric}
 \boldsymbol W=\boldsymbol F^{\mathsf H}
    \diag\{s_{W,k}^{2}\}\boldsymbol F,
\end{equation}
where $\boldsymbol F$ is the unitary discrete Fourier transform, or DFT, matrix and $\boldsymbol s_W$ is a fixed spectral-precision vector matched to the bandlimited GPS L1 C/A pulse shape across all branches, bit depths, and geometries. With $\boldsymbol C_\theta=\boldsymbol C(\boldsymbol\theta)$ and inter-satellite Gram matrix $\boldsymbol G_\theta=\boldsymbol C_\theta^{\mathsf H}\boldsymbol W\boldsymbol C_\theta$, profiling out complex satellite amplitudes yields
\begin{align}
 \widehat{\boldsymbol\alpha}(\boldsymbol\theta,\boldsymbol z)
 &=\boldsymbol G_\theta^{-1}\boldsymbol C_\theta^{\mathsf H}\boldsymbol W\boldsymbol z,\label{eq:amp}\\
 \boldsymbol B_\theta
 &=\boldsymbol W\boldsymbol C_\theta\boldsymbol G_\theta^{-1}
       \boldsymbol C_\theta^{\mathsf H}\boldsymbol W.\label{eq:B}
\end{align}
Each active branch estimates its state via bounded multistart maximization of the profiled quadratic contrast
\begin{equation}\label{eq:profile}
 \widehat{\boldsymbol\theta}_j\simeq
 \argmax_{\boldsymbol\theta\in\Theta}
 q(\boldsymbol\theta,\boldsymbol z_j),
 \qquad q(\boldsymbol\theta,\boldsymbol z)=\boldsymbol z^{\mathsf H}\boldsymbol B_\theta\boldsymbol z,
\end{equation}
equivalent to minimizing the $\boldsymbol W$-weighted waveform residual because $\boldsymbol z^{\mathsf H}\boldsymbol W\boldsymbol z$ is independent of $\boldsymbol\theta$. AGC scaling, quantization, and suppression act on all $N$ samples before projecting $\boldsymbol z_j$ onto a whitened Chebyshev replica subspace; when this subspace spans the replica manifold over $\Theta$, orthogonal energy adds only a state-independent constant to the residual norm.

\subsection{Risk Under a Processing Budget}
Active branch $j$ outputs the four-dimensional state $\widehat{\boldsymbol\theta}_j=[\widehat{\boldsymbol p}_j^{\mathsf T},\widehat b_{c,j}]^{\mathsf T}$ with position $\widehat{\boldsymbol p}_j=[\boldsymbol I_3\ \boldsymbol0]\widehat{\boldsymbol\theta}_j$. Rather than calibrating separate clock weights, the risk objective minimizes expected squared three-dimensional position error under a cap $S$ on active DPE solves per snapshot. Observing $\boldsymbol y$, an allocation policy activates at most $S$ front ends and forms convex simplex combinations along east, north, and up axes, unifying single-branch routing and sparse multibranch fusion; when auxiliary receiver clock-bias and four-dimensional search-boundary diagnostics are evaluated, active branch clock estimates are combined using the coordinate-averaged simplex weights $\widehat b_c=\sum_{j\in\mathcal A}\bar w_j\widehat b_{c,j}$ with $\bar w_j=\frac{1}{3}\sum_{d\in\{E,N,U\}}\widetilde w_{j,d}$.

\section{Sparse Risk-Calibrated Processing}\label{sec:method}
\method{} couples risk-calibrated sparse branch allocation with same-snapshot applicability gating. A scale-invariant waveform descriptor extracted from $\boldsymbol y$ indexes one of $M$ calibrated error models that either executes one matched front end or fuses up to $S=4$ branches, while an ordered reliability gate routes unfamiliar waveforms toward protective single-branch processors.

\subsection{A Fixed Library of Complementary Processors}
The core library comprises $J=8$ deterministic front ends: Branch 1 is pass-through Identity; Branch 2 applies radial time-domain Huber clipping at threshold $0.8$; Branch 3 cascades radial frequency Huber clipping at $2.5$ into radial time Huber clipping at $0.8$; Branch 4 cascades median-absolute-deviation, or MAD, frequency-to-time Huber clipping at $0.8$ in both stages; Branches 5 and 6 pair local widely linear I/Q weighting with $128$-point and $256$-point short-time Fourier transform, or STFT, attenuation; Branch 7 projects out up to four coherent spectral lines; and Branch 8 removes up to eight coherent lines followed by global spectral shrinkage.

For complex vector $\boldsymbol a$ with radial scale $\widehat\sigma(\boldsymbol a)=\max\{\operatorname{median}_n|a_n|/\sqrt{\ln2},10^{-12}\}$, radial Huber clipping acts elementwise as
\begin{equation}\label{eq:huber}
 [\mathcal H_\lambda(\boldsymbol a)]_n
 =a_n\min\{1,\lambda\widehat\sigma(\boldsymbol a)/|a_n|\},
\end{equation}
mapping zeros to zero. Frequency stages apply $\mathcal H_\lambda$ to $\boldsymbol F\boldsymbol y$, and cascades recompute scale at each stage. The MAD variant estimates centered real and imaginary scales separately in complex-RMS units, falling back to radial scaling whenever coarse quantization collapses either component median.

To suppress noncircular pulsed or modulated jammers, the local I/Q stage forms $\boldsymbol u_n=[\Re y_n,\Im y_n]^{\mathsf T}$ and computes its sample second-moment matrix over a reflected $257$-sample sliding window $\mathcal N_n$:
\begin{equation}\label{eq:localiq}
 \boldsymbol R_n=257^{-1}\sum_{m\in\mathcal N_n}\boldsymbol u_m\boldsymbol u_m^{\mathsf T}
 =\boldsymbol U_n\diag(\lambda_{n,1},\lambda_{n,2})\boldsymbol U_n^{\mathsf T}.
\end{equation}
Relative to the median minor eigenvalue $\nu=\max\{\operatorname{median}_n\min_k\lambda_{n,k},10^{-15}\}$, dominant directional energy is attenuated via
\begin{equation}\label{eq:wl}
 \widetilde{\boldsymbol u}_n=\boldsymbol U_n
 \diag\!\left\{\min\!\left(1,\frac{1.8\nu}{\max(\lambda_{n,k},10^{-15})}\right)\right\}
 \boldsymbol U_n^{\mathsf T}\boldsymbol u_n,
\end{equation}
implementing an adaptive widely linear filter $a_n y_n+c_n y_n^*$ \cite{widely,improper}. Subsequent STFT stages apply a periodic Hann window $h$ of length $L\in\{128,256\}$ and hop $H=L/4$,
\begin{equation}\label{eq:stft}
 X_{t,k}=L^{-1/2}\sum_{n=0}^{L-1}h_n y_{tH+n}\exp(-\mathrm i2\pi kn/L),
\end{equation}
shrinking bins against $\widehat\sigma_t=\operatorname{median}_k|X_{t,k}|/\sqrt{\ln2}$ via
\begin{equation}\label{eq:stftgain}
 \widetilde X_{t,k}=X_{t,k}\min\{1,(2.5\widehat\sigma_t/|X_{t,k}|)^2\},
\end{equation}
before dual-Hann overlap-add synthesis normalized pointwise by squared-window sums with reflection padding \cite{stft}. Coherent line processors detect spectral peaks satisfying
\begin{equation}\label{eq:linegate}
 \max_k|[\boldsymbol F\boldsymbol y]_k|^2>
 \widehat v\ln(N/10^{-3}),\quad
 \widehat v=\operatorname{median}_k|[\boldsymbol F\boldsymbol y]_k|^2/\ln2,
\end{equation}
initializing each tone on a twice-oversampled DFT grid, refining via three curvature-checked Newton steps bounded by a quarter bin, and projecting out stacked steering vectors $\boldsymbol A$ via
\begin{equation}\label{eq:lineproject}
 \boldsymbol r=\boldsymbol y-\boldsymbol A(\boldsymbol A^{\mathsf H}\boldsymbol A)^\dagger
                    \boldsymbol A^{\mathsf H}\boldsymbol y,
\end{equation}
until \eqref{eq:linegate} terminates or the four- or eight-line cap is reached, with Branch 8 appending spectral shrinkage of threshold $2.5$ and exponent $2$.

\subsection{Conditional Cross-Risk and Sparse Allocation}
Online bin assignment extracts the ten-dimensional descriptor $\boldsymbol f(\boldsymbol y)$ in Appendix~\ref{app:descriptor}, standardizes $\widetilde{\boldsymbol f}(\boldsymbol y)=\boldsymbol S^{-1}(\boldsymbol f(\boldsymbol y)-\boldsymbol\mu)$ using marginal mean $\boldsymbol\mu$ and diagonal standard deviation $\boldsymbol S$, and matches against $M=7$ centroids $\{\boldsymbol c_m\}_{m=0}^{M-1}$:
\begin{equation}\label{eq:descriptor}
 \mathsf B(\boldsymbol y)=\argmin_{m\in\{0,\ldots,M-1\}}
 \|\widetilde{\boldsymbol f}(\boldsymbol y)-\boldsymbol c_m\|_2^2,
\end{equation}
breaking ties toward the smallest index and recording $d_c^2=\|\widetilde{\boldsymbol f}(\boldsymbol y)-\boldsymbol c_{\mathsf B(\boldsymbol y)}\|_2^2$. Within bin $b$, branch error $\boldsymbol e_j=\widehat{\boldsymbol p}_j-\boldsymbol p_\star$ along coordinate $d\in\{E,N,U\}$ has conditional uncentered second-moment matrix
\begin{align}\label{eq:crossmoment}
 [\boldsymbol K_{b,d}]_{jk}
 &=\E[e_{j,d}e_{k,d}\mid\mathsf B=b]\\
 &=\Cov(e_{j,d},e_{k,d}\mid\mathsf B=b)+\beta_{j,b,d}\beta_{k,b,d},\nonumber
\end{align}
with bias $\beta_{j,b,d}=\E[e_{j,d}\mid\mathsf B=b]$. Simplex coordinate fusion
\begin{equation}\label{eq:fusion}
 \widehat p_d=\sum_j w_{b,j,d}\widehat p_{j,d},\quad
 \boldsymbol w_{b,d}\in\Delta_J=
 \{\boldsymbol w\geq\boldsymbol0:\boldsymbol1^{\mathsf T}\boldsymbol w=1\},
\end{equation}
attains exact conditional mean-squared position risk
\begin{equation}\label{eq:crossrisk}
 R_b(\boldsymbol w)=\E[\norm{\widehat{\boldsymbol p}-\boldsymbol p_\star}^2\mid\mathsf B=b]
 =\sum_d\boldsymbol w_{b,d}^{\mathsf T}\boldsymbol K_{b,d}\boldsymbol w_{b,d}.
\end{equation}
Because one DPE execution yields all three coordinates of $\widehat{\boldsymbol p}_j$, solver cost follows the joint support union
\begin{equation}\label{eq:support}
 \mathcal A_b=\bigcup_d\supp(\boldsymbol w_{b,d}),\qquad |\mathcal A_b|\leq S,
\end{equation}
with nominal cap $S=4$ and singleton vertices representing single-branch selection.

\subsection{Offline Risk Calibration and Validation Guard}
Preserving conditional independence during moment estimation requires fixing the representation prior to risk fitting. Descriptor definitions, candidate operators, gate thresholds, standardization moments $(\boldsymbol\mu,\boldsymbol S)$, centroids $\{\boldsymbol c_m\}_{m=0}^{M-1}$, and support threshold $\tau_d$ form the pre-frozen set $\mathcal D_{\mathrm{rep}}$, calibrated on an upstream development waveform stream before generating risk-calibration or test data. Conditional on $\mathcal D_{\mathrm{rep}}$, offline policy calibration uses two disjoint known-position splits of $\FitTrialCount{}$ risk-fit snapshots and $\GuardTrialCount{}$ validation-guard snapshots, where each snapshot independently draws its condition-state label with equal $1/14$ probability from the fourteen-cell core mixture rather than enforcing fixed per-cell quotas. Evaluating all $J=8$ branches on each quantized snapshot $\boldsymbol y_i$ preserves the operational joint error law, yielding empirical cross moments
\begin{equation}\label{eq:empK}
 \widehat{\boldsymbol K}_{b,d}=n_b^{-1}
   \sum_{i:\mathsf B_i=b}\boldsymbol e_{i,d}\boldsymbol e_{i,d}^{\mathsf T},
 \quad \boldsymbol e_{i,d}=[e_{i,1,d},\ldots,e_{i,J,d}]^{\mathsf T}.
\end{equation}
Enumerating all $\sum_{s=1}^4\binom8s=162$ nonempty supports decouples cardinality selection from convex optimization of
\begin{equation}\label{eq:weightopt}
 \widehat R_b(\boldsymbol w)=\sum_d\boldsymbol w_{b,d}^{\mathsf T}
       \widehat{\boldsymbol K}_{b,d}\boldsymbol w_{b,d},
\end{equation}
while the minimum-risk single branch on $\mathcal S_{\mathrm{fit}}$ defines each bin's anchor. On the independent validation split $\mathcal S_{\mathrm{val}}$, paired squared-error differences $D_i=\|\boldsymbol e_{\mathrm{fusion},i}\|^2-\|\boldsymbol e_{\mathrm{single},i}\|^2$ screen candidate supports via the one-sided normal standard-error rule
\begin{equation}\label{eq:calguard}
 \overline D+1.645\,\widehat{\operatorname{SE}}(\overline D)<0,
\end{equation}
requiring at least two validation snapshots in bin $b$ and reverting otherwise to the single-branch anchor. Criterion \eqref{eq:calguard} is a heuristic validation screen rather than a distribution-free finite-sample guarantee, since rare heavy-tailed gross errors may go unobserved in moderate validation splits and cause $\widehat{\operatorname{SE}}(\overline D)$ to understate uncertainty; variance-penalized empirical Bernstein bounds \cite{maurer2009empirical} over bounded error ranges $|D_i|\leq\sum_d E_d^2$ maintain valid finite-sample coverage even at small sample sizes, yet remain too conservative to certify moderate risk reductions without substantially larger validation splits. Selected supports and coordinate weights remain frozen across all evaluations.

\subsection{Three-Stage Online Reliability Gating}
Online reliability gating, denoted RG, screens the current snapshot $\boldsymbol y$ using known bit depth $B$, five descriptor features $p=f_1$, $a_4=f_5$, $r=f_6$, $c_1=f_7$, and $\iota=f_9$, smoothed $65$-bin spectral percentiles $s_{90}$ and $s_{10}$, and the $64$-sample periodic-Hann concentration metric
\begin{equation}\label{eq:concentration}
 \chi=\operatorname{median}_t
 \frac{\max_k|X_{t,k}|^2}{\sum_k|X_{t,k}|^2},
\end{equation}
evaluated at hop $16$ against the pre-frozen thresholds in Table~\ref{tab:rgparams} of Appendix~\ref{app:descriptor}. Snapshots within descriptor support $d_c^2\leq\tau_d$ retain the frozen core policy immediately. Out-of-support snapshots with weak lag-one correlation $c_1<\tau_c$ enter the low-lag screen: proper transient bursts satisfying $r>\tau_r$ and $\iota<\tau_\iota$ apply a reflected $65$-sample local-power weighter relative to $\nu_p=\operatorname{quantile}_{\alpha_p}\{p_n\}$,
\begin{equation}\label{eq:localpower}
 z_n=y_n\min\{1,\gamma_p\nu_p/\max(p_n,10^{-20})\},
\end{equation}
whereas remaining low-lag inputs select Identity when $p<\tau_{p,\mathrm{low}}$ and four-line projection when $p\geq\tau_{p,\mathrm{low}}$. Snapshots not matched earlier retain the core policy if a narrowband peak satisfies $p>\tau_{p,\mathrm{high}}$ and $s_{90}<\tau_{s90}$; otherwise $1$-bit inputs route to unweighted $256$-point STFT attenuation, stationary $2$-bit inputs with $r<\tau_r$ and $a_4\geq\tau_{a4}$ route to frequency Huber clipping at $\lambda_{\mathrm{RG}}$ when $\chi<\tau_\chi$ or $64$-point STFT attenuation when $\chi\geq\tau_\chi$, higher-resolution inputs with $B>2$, $r<\tau_r$, $\chi\geq\tau_\chi$, and $s_{10}\geq\tau_{s10}$ likewise invoke $64$-point STFT attenuation, and remaining cases retain the core policy. Every protective route executes one DPE solve as summarized in Algorithm~\ref{alg:crisp}, with expected solve count $\E[L]=\sum_{b,a}\Pr(\mathsf B=b,A=a)L(b,a)\in[1,4]$.

\begin{algorithm}[t]
\caption{Online Single-Snapshot Sparse Allocation and Fusion}
\label{alg:crisp}\label{alg:receiver}
\footnotesize
\begin{algorithmic}[1]
\Require Quantized baseband snapshot $\boldsymbol y\in\C^N$; ADC bit depth $B$; frozen moments $(\boldsymbol\mu,\boldsymbol S)$, centroids $\{\boldsymbol c_m\}_{m=0}^{M-1}$, bin weights $\{\boldsymbol w_{m,d}\}$, frozen gate thresholds in Table~\ref{tab:rgparams}, spectral metric $\boldsymbol W$, and operators $\{\mathcal R_j\}$.
\Ensure Three-dimensional ENU position estimate $\widehat{\boldsymbol p}\in\R^3$.
\State Extract $\boldsymbol f(\boldsymbol y)$ via \eqref{eq:features} and standardize $\widetilde{\boldsymbol f}(\boldsymbol y)=\boldsymbol S^{-1}(\boldsymbol f(\boldsymbol y)-\boldsymbol\mu)$.
\State Assign bin $b=\argmin_{m}\|\widetilde{\boldsymbol f}(\boldsymbol y)-\boldsymbol c_m\|_2^2$ and set $d_c^2=\|\widetilde{\boldsymbol f}(\boldsymbol y)-\boldsymbol c_b\|_2^2$.
\State Compute smoothed spectral percentiles $s_{10},s_{90}$ and short-frame concentration $\chi=\operatorname{median}_t\{\max_k|X_{t,k}|^2/\sum_k|X_{t,k}|^2\}$.
\If{$d_c^2\leq\tau_d$ \textbf{or} $(c_1\geq\tau_c\land p>\tau_{p,\mathrm{high}}\land s_{90}<\tau_{s90})$}
    \State Retain core policy: $\widetilde{\boldsymbol w}_d\leftarrow\boldsymbol w_{b,d}$ for $d\in\{E,N,U\}$.
\ElsIf{$c_1<\tau_c$}
    \State Select low-lag protective index $j_\star$ via $r>\tau_r\land\iota<\tau_\iota$ using \eqref{eq:localpower}, $p<\tau_{p,\mathrm{low}}$, or $p\geq\tau_{p,\mathrm{low}}$; set $\widetilde{\boldsymbol w}_d\leftarrow\boldsymbol e_{j_\star}$.
\Else
    \State Evaluate quantizer-occupancy predicates $q_5,q_6,q_7$ in \eqref{eq:gatepredicates} to assign protective unit vector $\boldsymbol e_{j_\star}$ or core weights $\boldsymbol w_{b,d}$.
\EndIf
\State Form active branch support $\mathcal A=\bigcup_{d\in\{E,N,U\}}\supp(\widetilde{\boldsymbol w}_d)$ with $|\mathcal A|\leq S$.
\For{each $j\in\mathcal A$}
    \State Suppress interference $\boldsymbol z_j=\mathcal R_j(\boldsymbol y)$ and project onto the replica subspace.
    \State Solve $\widehat{\boldsymbol\theta}_j=\argmax_{\boldsymbol\theta\in\Theta}\boldsymbol z_j^{\mathsf H}\boldsymbol W\boldsymbol C_\theta(\boldsymbol C_\theta^{\mathsf H}\boldsymbol W\boldsymbol C_\theta)^{-1}\boldsymbol C_\theta^{\mathsf H}\boldsymbol W\boldsymbol z_j$.
    \State Extract branch position $\widehat{\boldsymbol p}_j=[\boldsymbol I_3\ \boldsymbol0]\widehat{\boldsymbol\theta}_j$.
\EndFor
\If{$|\mathcal A|=1$}
    \State \Return single-branch estimate $\widehat{\boldsymbol p}\leftarrow\widehat{\boldsymbol p}_{j_\star}$ for $\mathcal A=\{j_\star\}$.
\Else
    \State Fuse $\widehat p_d=\sum_{j\in\mathcal A}\widetilde w_{j,d}\widehat p_{j,d}$ for $d\in\{E,N,U\}$ and \Return $\widehat{\boldsymbol p}$.
\EndIf
\end{algorithmic}
\end{algorithm}

\section{Risk Interpretation and Finite Calibration}\label{sec:theory}
\subsection{Position Error After Nonlinear Processing}
Conditioned on a fixed navigation state, environment, and routing event of positive probability, let $\boldsymbol m=\E[\boldsymbol z]$ and $\boldsymbol\Sigma=\Cov(\boldsymbol z)$ denote the post-suppression mean waveform and covariance. Taking the expectation of \eqref{eq:profile} yields
\begin{equation}\label{eq:population}
 \overline q(\boldsymbol\theta)
 =\boldsymbol m^{\mathsf H}\boldsymbol B_{\theta}\boldsymbol m
  +\tr(\boldsymbol B_{\theta}\boldsymbol\Sigma),
\end{equation}
separating coherent mean-waveform energy from residual noise and interference projected onto the replica subspace.

\begin{lemma}[Profiling and expectation]\label{prop:profile}
If $\boldsymbol G_\theta$ is nonsingular and $\boldsymbol z$ has a finite second moment, then
\begin{align}
 &\min_{\boldsymbol\alpha}\E\norm{\boldsymbol z-\boldsymbol C_\theta\boldsymbol\alpha}_{W}^{2}
  -\E\min_{\boldsymbol\alpha}\norm{\boldsymbol z-\boldsymbol C_\theta\boldsymbol\alpha}_{W}^{2}
   \nonumber\\
 &\hspace{25mm}=\tr(\boldsymbol B_\theta\boldsymbol\Sigma)\geq0.\label{eq:interchange}
\end{align}
\end{lemma}
\begin{proof}
Expanding the quadratic forms gives $\boldsymbol m^{\mathsf H}(\boldsymbol W-\boldsymbol B_\theta)\boldsymbol m+\tr(\boldsymbol W\boldsymbol\Sigma)$ and $\boldsymbol m^{\mathsf H}(\boldsymbol W-\boldsymbol B_\theta)\boldsymbol m+\tr[(\boldsymbol W-\boldsymbol B_\theta)\boldsymbol\Sigma]$, whose difference $\tr(\boldsymbol B_\theta\boldsymbol\Sigma)$ is nonnegative for positive semidefinite $\boldsymbol B_\theta$ and $\boldsymbol\Sigma$.
\end{proof}

When $\boldsymbol\Sigma=\sigma^2\boldsymbol W^{-1}$, the trace equals the constant $\sigma^2K$. Quantization and nonlinear clipping break proportionality to $\boldsymbol W^{-1}$. With $\overline{\boldsymbol\alpha}=\boldsymbol G_\theta^{-1}\boldsymbol C_\theta^{\mathsf H}\boldsymbol W\boldsymbol m$ and $\boldsymbol P_\theta^\perp=\boldsymbol I-\boldsymbol C_\theta\boldsymbol G_\theta^{-1}\boldsymbol C_\theta^{\mathsf H}\boldsymbol W$, differentiating \eqref{eq:population} with respect to $\theta_a$ yields
\begin{align}\label{eq:popgradient}
 \frac{\partial\overline q}{\partial\theta_a}
 &=2\Re\{(\boldsymbol C_{\theta,a}\overline{\boldsymbol\alpha})^{\mathsf H}
            \boldsymbol W\boldsymbol P_\theta^\perp\boldsymbol m\}
   +\tr(\boldsymbol B_{\theta,a}\boldsymbol\Sigma),
\end{align}
showing how nonlinear processing shifts the contrast maximizer via both mean-waveform distortion $\boldsymbol m$ and state-dependent subspace covariance $\tr(\boldsymbol B_{\theta}\boldsymbol\Sigma)$ \cite{mismatch,scatter,delaymismatch}. Around a regular interior maximizer $\boldsymbol\theta_j^{\mathrm P}$ with $\boldsymbol A_j=-\nabla^2\overline q_j(\boldsymbol\theta_j^{\mathrm P})\succ0$ and centered score $\boldsymbol s_j=\nabla q_j(\boldsymbol\theta_j^{\mathrm P},\boldsymbol z_j)$, first-order expansion with $\boldsymbol P=[\boldsymbol I_3\ \boldsymbol0]$ gives
\begin{equation}\label{eq:scoreerror}
 \boldsymbol e_j\simeq\boldsymbol\delta_j+\boldsymbol L_j\boldsymbol s_j,
 \quad \boldsymbol\delta_j=\boldsymbol P(\boldsymbol\theta_j^{\mathrm P}-\boldsymbol\theta_\star),
 \quad \boldsymbol L_j=\boldsymbol P\boldsymbol A_j^{-1},
\end{equation}
and inner-product cross moment
\begin{equation}\label{eq:crossscore}
 \E[\boldsymbol e_j^{\mathsf T}\boldsymbol e_k]
 \simeq\boldsymbol\delta_j^{\mathsf T}\boldsymbol\delta_k
 +\tr\{\boldsymbol L_j\Cov(\boldsymbol s_j,\boldsymbol s_k)\boldsymbol L_k^{\mathsf T}\}.
\end{equation}
Appendix~\ref{app:quadratic} gives the exact fourth-cumulant score covariance for non-Gaussian quantized waveforms, while empirical paired-error calibration captures both local score coupling and nonlocal sidelobe events on the shared snapshot.

\subsection{Why Joint Error Moments Matter}
Because the support-constrained coordinate-simplex class $\mathcal W_S$ contains every common-vertex single-branch policy, population risk inherits the simplex-stacking inclusion bound \cite{breiman1996stacked}:
\begin{equation}\label{eq:inclusion}
 \min_{\boldsymbol w\in\mathcal W_S}R_b(\boldsymbol w)
 \leq\min_{j\leq J}\E[\norm{\boldsymbol e_j}^2\mid\mathsf B=b].
\end{equation}
Combining two branches $j$ and $k$ with weight $t\in[0,1]$ on branch $k$ along one coordinate gives \cite{breiman1996stacked}
\begin{align}\label{eq:direction}
 R(t)&=K_{jj}+2t(K_{jk}-K_{jj})\\[-1mm]
 &\quad+t^2(K_{jj}+K_{kk}-2K_{jk}).\nonumber
\end{align}
Whenever $K_{jk}<K_{jj}$, the slope at $t=0$ is strictly negative, so an auxiliary branch with $K_{kk}>K_{jj}$ still lowers fused risk when its residual distortion is weakly or negatively correlated with the anchor. Uncentered moments $K_{jk}=\Cov(e_j,e_k)+\E[e_j]\E[e_k]$ evaluate exact mean-squared risk of uncorrected branch outputs directly.

\subsection{A Finite-Sample Calibration Bound}
Let $D=3$ be the spatial dimension, $M$ the bin count, and $\mathcal D_{\mathrm{rep}}$ collect all upstream-fitted quantities, comprising $(\boldsymbol\mu,\boldsymbol S)$, $\{\boldsymbol c_m\}_{m=0}^{M-1}$, support threshold $\tau_d$, the operator library, shared spectral metric $\boldsymbol W$, and bounded search domain $\Theta$, all fixed on an upstream set $\mathcal S_{\mathrm{rep}}$ prior to risk fitting. Let $\mathcal S_{\mathrm{fit}}=\{(\boldsymbol y_i,\boldsymbol\theta_{\star,i})\}_{i=1}^{N_{\mathrm{fit}}}$ be drawn i.i.d.\ from the target mixture law independently of $\mathcal D_{\mathrm{rep}}$. Because $\mathsf B_i=\mathsf B(\boldsymbol y_i;\mathcal D_{\mathrm{rep}})$ depends only on snapshot $i$ once $\mathcal D_{\mathrm{rep}}$ is fixed, conditioning on $\mathcal F_{\mathrm{bin}}=\sigma(\mathcal D_{\mathrm{rep}},\mathsf B_1,\ldots,\mathsf B_{N_{\mathrm{fit}}})$ preserves mutual conditional independence across snapshots: for any bin $b$ with $n_b=\sum_{i=1}^{N_{\mathrm{fit}}}\boldsymbol1\{\mathsf B_i=b\}>0$, the $n_b$ error matrices $\{\boldsymbol e_i:\mathsf B_i=b\}$ are conditionally i.i.d.\ draws from $\mathbb P(\boldsymbol e\in\cdot\mid\mathsf B(\boldsymbol y;\mathcal D_{\mathrm{rep}})=b,\mathcal D_{\mathrm{rep}})$ while allowing arbitrary inter-branch and inter-coordinate correlation within each snapshot, whereas fitting $(\boldsymbol\mu,\boldsymbol S)$ or $\{\boldsymbol c_m\}$ on $\mathcal S_{\mathrm{fit}}$ would couple $\mathsf B_i$ across snapshots. With solver bounds $[\ell_d,u_d]$ and truth bounds $[a_d,c_d]$, branch errors satisfy
\begin{equation}\label{eq:errorenvelope}
 |e_{j,d}|\leq E_d,
 \qquad E_d=\max\{|\ell_d-c_d|,|u_d-a_d|\},
\end{equation}
giving $E_d=600$ m for $[-300,300]$-m bounds or tighter constants over narrower prior cubes.

\begin{theorem}[Simultaneous calibration accuracy]\label{prop:calibration}
For $\mathcal S_{\mathrm{fit}}$ drawn i.i.d.\ independently of $\mathcal D_{\mathrm{rep}}$, conditional on $\mathcal F_{\mathrm{bin}}=\sigma(\mathcal D_{\mathrm{rep}},\mathsf B_1,\ldots,\mathsf B_{N_{\mathrm{fit}}})$ on any event where $n_b>0$ for observed bins, define for any $0<\delta<1$
\begin{equation}\label{eq:samplebound}
 \epsilon_{b,d}=E_d^2
  \sqrt{\frac{2\log(2DMJ^2/\delta)}{n_b}}.
\end{equation}
With conditional probability at least $1-\delta$, simultaneously across all observed bins and coordinates,
\begin{align}
 \norm{\widehat{\boldsymbol K}_{b,d}-\boldsymbol K_{b,d}}_{\max}
 &\leq\epsilon_{b,d},\label{eq:momentconcentration}\\
 \sup_{\boldsymbol w\in\mathcal W_S}
 |\widehat R_b(\boldsymbol w)-R_b(\boldsymbol w)|
 &\leq\sum_d\epsilon_{b,d},\label{eq:uniform}
\end{align}
where $\|\cdot\|_{\max}$ denotes the maximum absolute matrix entry.
\end{theorem}
\begin{proof}
Given $\mathcal F_{\mathrm{bin}}$, the index set $I_b=\{i:\mathsf B_i=b\}$ of size $n_b$ is $\mathcal F_{\mathrm{bin}}$-measurable, and mutual independence of $(\boldsymbol y_i,\boldsymbol\theta_{\star,i})$ given $\mathcal D_{\mathrm{rep}}$ renders $\{e_{i,j,d}e_{i,k,d}\}_{i\in I_b}$ conditionally independent across $i\in I_b$ with conditional mean $[\boldsymbol K_{b,d}]_{jk}$ and range $[-E_d^2,E_d^2]$. Hoeffding's inequality \cite{hoeffding1963} and a union bound over at most $DMJ^2$ entries yield \eqref{eq:momentconcentration}. For $\boldsymbol w\in\Delta_J$, $|\boldsymbol w^{\mathsf T}\boldsymbol H\boldsymbol w|\leq\|\boldsymbol H\|_{\max}$, which summed over $d\in\{E,N,U\}$ gives \eqref{eq:uniform}.
\end{proof}

\begin{corollary}[Optimization and conservative-selection costs]\label{cor:excess}
Suppose weights are fitted to $\widetilde{\boldsymbol K}_{b,d}$ satisfying $\|\widetilde{\boldsymbol K}_{b,d}-\widehat{\boldsymbol K}_{b,d}\|_{\max}\leq\rho_{b,d}$, producing $\widehat{\boldsymbol w}_b$ with suboptimality gap at most $\eta_b$ on $\widetilde R_b$, and validation selection $\overline{\boldsymbol w}_b$ satisfies
\begin{equation}\label{eq:guardcost}
 \widetilde R_b(\overline{\boldsymbol w}_b)
 \leq\widetilde R_b(\widehat{\boldsymbol w}_b)+\gamma_b.
\end{equation}
On the high-probability event of Theorem~\ref{prop:calibration},
\begin{align}\label{eq:oracle}
 R_b(\overline{\boldsymbol w}_b)-\min_{\boldsymbol w\in\mathcal W_S}R_b(\boldsymbol w)
 \leq2\sum_d(\epsilon_{b,d}+\rho_{b,d})+\eta_b+\gamma_b.
\end{align}
\end{corollary}
\begin{proof}
Combining \eqref{eq:uniform} with $\|\widetilde{\boldsymbol K}_{b,d}-\widehat{\boldsymbol K}_{b,d}\|_{\max}\leq\rho_{b,d}$ bounds $|\widetilde R_b(\boldsymbol w)-R_b(\boldsymbol w)|\leq\sum_d(\epsilon_{b,d}+\rho_{b,d})$ uniformly on $\mathcal W_S$; evaluating at $\overline{\boldsymbol w}_b$ and the population minimizer with \eqref{eq:guardcost} yields \eqref{eq:oracle}.
\end{proof}

Corollary~\ref{cor:excess} decomposes excess risk into bin concentration $\epsilon_{b,d}\propto n_b^{-1/2}$, matrix shrinkage $\rho_{b,d}$ \cite{shrink}, convex Frank-Wolfe duality gap $\eta_b$ certified on each enumerated support subproblem, and validation sparsity trade-off $\gamma_b=[\widetilde R_b(\overline{\boldsymbol w}_b)-\widetilde R_b(\widehat{\boldsymbol w}_b)]_+$.

\subsection{Same-Observation Gating and Coordinate Geometry}\label{sec:gatinggeom}
When online gating reroutes snapshot $\boldsymbol y$ on event $A$, replacing core error $\boldsymbol e_{\mathrm C}$ with protective error $\boldsymbol e_{\mathrm R}$, partitioning over $A$ and $A^{\mathrm c}$ yields the exact population risk identity
\begin{equation}\label{eq:gaterisk}
 R_{\mathrm{RG}}-R_{\mathrm C}
 =\E\!\left[\boldsymbol1_A
   (\norm{\boldsymbol e_{\mathrm R}}^2-\norm{\boldsymbol e_{\mathrm C}}^2)\right],
\end{equation}
retaining same-snapshot statistical dependence between gate predicates and DPE outputs. Identity \eqref{eq:gaterisk} does not guarantee non-degradation: net risk change follows the expectation of $\norm{\boldsymbol e_{\mathrm R}}^2-\norm{\boldsymbol e_{\mathrm C}}^2$ over rerouted operating regimes. For fixed weights $\{\boldsymbol w_{b,d}\}$, simplex fusion $\widehat p_d=\sum_j w_{b,j,d}\widehat p_{j,d}$ is translation equivariant to common branch shifts $\boldsymbol t\in\R^3$ because $\boldsymbol1^{\mathsf T}\boldsymbol w_{b,d}=1$ and stays within the coordinatewise convex hull of active branches. Physical receiver displacement additionally alters line-of-sight vectors, sub-sample code phases, and search-box margins, motivating empirical translated-state and aiding-range tests in Section~\ref{sec:results}. Distinct diagonal matrices $\boldsymbol D_j$ in $\widehat{\boldsymbol p}=\sum_j\boldsymbol D_j\widehat{\boldsymbol p}_j$ provide intentional ENU anisotropy matched to directional dilution of precision.

\section{Experimental Protocol}\label{sec:protocol}
All evaluations use controlled Monte Carlo baseband simulations in which every receiver processes identical $2$-ms quantized snapshots under the shared spectral metric $\boldsymbol W$, identical satellite replicas, and identical per-branch multistart DPE budgets. Independent pseudorandom streams separate upstream representation development, baseline calibration, risk fitting, validation screening, the $\PaperObservationCount$-snapshot primary and transfer benchmark, and the standalone aiding, directional-bias, and serial-timing diagnostic suites summarized in Table~\ref{tab:provenance}; all primary, geometry-transfer, matched-ablation, and gating results in Tables~\ref{tab:main}--\ref{tab:rg} evaluate the pre-frozen representation $\mathcal D_{\mathrm{rep}}$ across these strictly disjoint splits.

\begin{table*}[!t]
\centering\footnotesize
\caption{Dataset Splits, Operating Regimes, Calibrated Quantities, and Frozen Parameters}
\label{tab:provenance}
\setlength{\tabcolsep}{3.5pt}
\begin{tabularx}{\textwidth}{@{}>{\looseness=0\raggedright\arraybackslash}p{2.9cm}>{\looseness=0\raggedright\arraybackslash}p{2.4cm}>{\looseness=0\raggedright\arraybackslash}p{3.9cm}>{\looseness=0\raggedright\arraybackslash}p{4.3cm}>{\looseness=0\raggedright\arraybackslash}X@{}}
\toprule
Stage & Snapshots & Operating regime and truth range & Calibrated quantities & Strictly frozen quantities \\
\midrule
Upstream representation development $\mathcal S_{\mathrm{rep}}$ & $1\,792$ development waveforms & G0 at PDOP $2.43$; fixed $\boldsymbol\theta_{\mathrm{ref}}$ and $\pm60$-m pos., $\pm30$-m clk. & Spectral $\boldsymbol s_W$, moments $(\boldsymbol\mu,\boldsymbol S)$, centroids $\{\boldsymbol c_m\}_{m=0}^6$, gate thresholds in Table~\ref{tab:rgparams} & Operator bank $\mathcal R_1$--$\mathcal R_8$ and search domain $[-300,300]^4$ m \\
\addlinespace[1.2pt]
Baseline calibration & $3\,584$ RIM and clean FrFT streams & G0 across $14$ core cells at $256$ per cell, fixed and $\pm60/\pm30$ m & One global $\lambda\in\{0.8,1.345,2.5\}$ per RIM family; bit-matched Weibull $\gamma_B$ at $P_{\mathrm{FA}}=10^{-7}$ for PLSE-FrFT & Shared metric $\boldsymbol W$, Chebyshev replica basis, multistart DPE solver \\
\addlinespace[1.2pt]
Disjoint representation audit & $\RepFitTrialCount{}$ label-free waveforms & G0 across $14$ core cells at $64$ per cell without state labels & Refitted $(\widehat{\boldsymbol\mu},\widehat{\boldsymbol S})$ and $k$-means $\{\widehat{\boldsymbol c}_m\}_{m=0}^6$ & Downstream $\mathcal S_{\mathrm{fit}}$, $\mathcal S_{\mathrm{val}}$, and $9\,856$ test waveforms \\
\addlinespace[1.2pt]
Cross-risk moment fit $\mathcal S_{\mathrm{fit}}$ & $\FitTrialCount{}$ i.i.d.\ mixture draws & G0 uniform $14$-cell mixture, $109$--$142$ per cell, fixed and $\pm60/\pm30$ m & Empirical moments $\widehat{\boldsymbol K}_{b,d}$, candidate weights $\widehat{\boldsymbol w}_b$, single anchors, continuous routers & Pre-frozen $\mathcal D_{\mathrm{rep}}$, operators $\mathcal R_1$--$\mathcal R_8$, metric $\boldsymbol W$, domain $\Theta$ \\
\addlinespace[1.2pt]
Validation screen $\mathcal S_{\mathrm{val}}$ & $\GuardTrialCount{}$ i.i.d.\ mixture draws & G0 uniform $14$-cell mixture, $49$--$82$ per cell, fixed and $\pm60/\pm30$ m & Binary retention via \eqref{eq:calguard} between $\widehat{\boldsymbol w}_b$ and bin anchor $\boldsymbol e_{j_\star}$ & Pre-frozen $\mathcal D_{\mathrm{rep}}$, candidate weights $\widehat{\boldsymbol w}_b$, single anchors \\
\addlinespace[1.2pt]
Primary and transfer benchmarks & $\PaperObservationCount{}$ stratified across $201$ cells & Fixed cell quotas on G0--G5; fixed, $\pm60/\pm120/\pm180$ m, shifted centers, hopping & None & All CRISP weights, gate thresholds, and five baseline settings \\
\addlinespace[1.2pt]
Aiding, bias, and timing suites & $4\,608$ aid, $672$ bias, $56{+}304$ timed & G0 across $18$ aiding points; G3 in-phase multipath & None on aiding or timing; $288$-trial local moment fit and $384$-trial disjoint test on bias diagnostic & Frozen primary CRISP and baseline parameters \\
\bottomrule
\end{tabularx}
\end{table*}

\subsection{Signal, Geometry, and Search Settings}
Six GPS satellites transmit L1 C/A PRN codes $1$ through $6$ over geometric ranges of $20.20$ to $23.40$ Mm. Complex baseband sampling at $f_s=4.092$ MHz yields four samples per chip and $N=8,184$ complex samples per $2$-ms snapshot, shaped by a flat passband up to $0.85$ chip rate with raised-cosine roll-off to $1.15$ chip rate. Nominal settings use $C/N_0=46$ dB-Hz, unit pre-AGC thermal noise variance, and $B=2$ bits per I/Q component. Fixed-state trials set $\boldsymbol\theta_{\mathrm{ref}}=[23,-31,14,37]^{\mathsf T}$ m, while displaced-state trials add uniform position offsets over $[-60,60]$ m per axis and clock offsets over $[-30,30]$ m. Each DPE solve searches $[-300,300]^4$ m from a coarse $3^4$ full-range grid, a refined $3^4$ inner grid at $\{-55,0,55\}$ m, and $32$ Sobol points, polishing the top two candidates via $55$ L-BFGS-B iterations using a degree-$64$ Chebyshev basis spanning $\pm3.1$ chips of rank $186$, waveform relative error $2.97\times10^{-12}$, and gradient relative error below $5.01\times10^{-9}$.

\begin{figure*}[!t]\centering
\subfloat[Satellite-angle layouts]{\includegraphics[width=.455\textwidth]{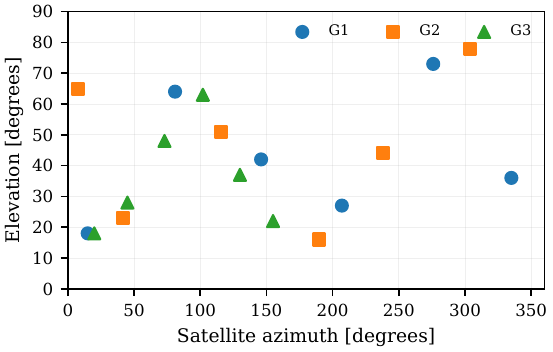}}\hspace{0.03\textwidth}%
\subfloat[Positioning RMSE]{\includegraphics[width=.455\textwidth]{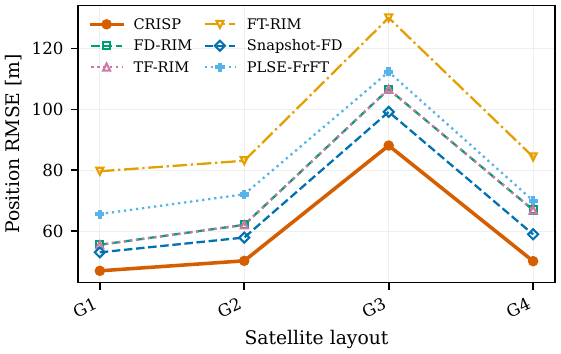}}
\caption{Satellite constellation layouts and positioning RMSE across changed geometries G1 through G4.}\label{fig:geometrycombined}
\end{figure*}

Six constellations G0 through G5 probe geometric conditioning, with representative skyplots in Fig.~\ref{fig:geometrycombined}(a). Nominal calibration layout G0 has PDOP $2.43$, while G1 and G2 span all azimuth quadrants at PDOPs $2.26$ and $2.38$. Layout G3 clusters all six satellites within $20^\circ$ to $155^\circ$ azimuth at PDOP $4.33$, G4 rotates G0 horizontally by $45^\circ$ at PDOP $2.43$, and G5 forms an unseen holdout geometry at PDOP $2.15$. Geometry changes recompute propagation delays and replica bases while keeping all receiver weights and thresholds fixed.

\subsection{Interference Regimes and Transfer Stress Suites}
The fourteen-cell core benchmark crosses fixed and displaced states with seven environments: Clean, continuous-wave jamming CW10 and CW20 at $10$ and $20$ dB J/N centered at $0.13f_s$ with $200$-Hz Gaussian frequency jitter, real Gaussian-envelope pulsed jamming Pulse10 and Pulse20, and compound jamming Mixed10 and Mixed20 splitting power $70\%$ into a gated chirp and $30\%$ into a real pulse. Offline calibration samples cell labels independently and uniformly with probability $1/14$ across this fourteen-cell mixture to yield $\FitTrialCount{}$ risk-fit snapshots at $109$ to $142$ trials per cell and $\GuardTrialCount{}$ validation snapshots at $49$ to $82$ trials per cell prior to test generation, whereas test suites use fixed stratified per-cell counts such as $384$ trials per cell across the $\CoreTrials{}$-snapshot core benchmark. Parametric transfer suites sweep $C/N_0\in\{36,40,46,50\}$ dB-Hz, bit depth $B\in\{1,2,4,8\}$, and $\text{J/N}\in[-10,40]$ dB, plus $36$ periodic $1$-MHz linear and quadratic frequency-modulated chirp settings governed on $u\in[0,1)$ by
\begin{align}
 f_{\mathrm{LFM}}(u)&=-B_c/2+B_cu,\\
 f_{\mathrm{QFM}}(u)&=-B_c/2+B_c(0.15u+0.85u^2),\label{eq:chirplaw}
\end{align}
with LFM periods $\{10,20,30\}\,\mu\text{s}$ and QFM periods $\{100,200,300\}\,\mu\text{s}$. Additional suites test $\HoldoutTrials{}$ G5 holdout trials, $1,152$ translated-center trials shifted by $[80,-40,30]^{\mathsf T}$, $[140,70,-50]^{\mathsf T}$, and $[-130,110,60]^{\mathsf T}$ m, truth half-widths of $120$ and $180$ m, specular multipath of $0.5$ relative power at $\{0.2,0.5,1.0\}$-chip delays, and $\HoppingTrials{}$ fast frequency-hopping trials at $20$ dB J/N with phase-continuous $20$ to $100\,\mu\text{s}$ dwells across $[-0.4f_s,0.4f_s]$.

\subsection{Baselines, Ablation Controls, and Statistical Evaluation}
Five published interference-mitigation baselines share the quantized $2$-ms input and DPE backend after independent development-set calibration. Full-block FD-RIM, TF-RIM, and FT-RIM \cite{rim} operate across all $N=8,184$ samples using centered I/Q MAD scales with radial-median fallback under low-bit collapse, selecting one global Huber threshold over $\lambda\in\{0.8,1.345,2.5\}$ by minimizing equal-cell 3-D MSE across an independent $3,584$-snapshot fourteen-cell calibration grid to yield frozen thresholds $1.345$, $1.345$, and $2.5$. Subblock Snapshot-FD \cite{snapshot} partitions $\boldsymbol y$ into eight nonoverlapping $0.25$-ms subblocks of $1,023$ samples with per-subblock I/Q MAD scaling and radial fallback, selecting $\lambda=1.345$ on the same $3,584$-snapshot grid. PLSE-FrFT \cite{plse} adapts piecewise FrFT chirp blanking \cite{frft,dfrft} to $2$-ms low-bit snapshots by extracting dominant ridges on a $16$-sample periodic-Hann STFT with $4$-sample hop and $128$-point DFT, segmenting at frequency jumps, fitting up to six generalized cross-validated linear pieces, rotating each piece via fast Ozaktas discrete FrFT \cite{frft} for signum blanking above a bit-matched Weibull threshold $\gamma_B$ calibrated on clean quantized transforms at $P_{\mathrm{FA}}=10^{-7}$, and applying residual synthesis $\boldsymbol y+\mathcal F_{-\hat a}\{\widetilde{\boldsymbol X}_{\hat a}-\boldsymbol X_{\hat a}\}$. Each active branch invokes the common multistart DPE solver, while total receiver runtime reflects both front-end transforms and active branch count.

Matched-library controls calibrated on the identical $\FitTrialCount{}$ risk-fit and $\GuardTrialCount{}$ validation sets isolate internal mechanisms: \textit{Single} selects each bin's minimum-MSE branch; \textit{RidgeRouter-Single} and \textit{kNNRouter-Single} train continuous $10$-D feature cost-sensitive single-branch routers via cross-validated quadratic ridge regression and distance-weighted $k$-nearest neighbors on $\mathcal S_{\mathrm{fit}}$; \textit{Diagonal} uses diagonal error moments only; \textit{CenteredCov} uses centered error covariances; \textit{ScalarGuarded} and \textit{ScalarFullK} solve isotropic scalar-weight simplex programs $\boldsymbol w^{\mathsf T}(\sum_d\widehat{\boldsymbol K}_{b,d})\boldsymbol w$ across ENU axes with and without validation pruning; \textit{FullK} optimizes coordinate-anisotropic uncentered moments without validation pruning or online gating; and \textit{Core} denotes the validation-pruned coordinate policy prior to online RG. Two static-prior diagnostics output $[23,-31,14]^{\mathsf T}$ m and the empirical risk-fit truth mean.

Accuracy is measured by three-dimensional root-mean-square error over all $n$ trials without outlier removal,
\begin{equation}\label{eq:rmse}
 \operatorname{RMSE}=\sqrt{n^{-1}\sum_i \norm{\widehat{\boldsymbol p}_i-\boldsymbol p_{\star,i}}_2^2},
\end{equation}
alongside receiver clock-bias RMSE, error percentiles P95 and P99, exceedance rate $P(e>100\,\mathrm{m})$, capture rate $P(e\leq 100\,\mathrm{m})$, and boundary hit rate $\Pr(\max_a|\widehat\theta_a|\geq 299.99\,\mathrm{m})$. Uncertainty is quantified via $20,000$ paired within-stratum bootstrap resamples using Bonferroni simultaneous $95\%$ confidence intervals across comparison families and paired marginal $95\%$ intervals for targeted contrasts. Simulations are implemented with the PyTorch 2.8.0, NumPy, and SciPy libraries in a Python 3.11 environment on a workstation equipped with a 12th Gen Intel Core i9-12900KF CPU and an NVIDIA GeForce RTX 5060 Ti 16 GB GPU supporting parallel complex-valued baseband computations, whereas single-threaded serial latency covering descriptor extraction, gating, active front ends, projection, DPE, and fusion is timed on the CPU using one BLAS thread after parallel workers stop, excluding signal synthesis, offline calibration, replica-cache initialization, and data input-output, across the $56$-snapshot core timing stream and an expanded $304$-snapshot stratified suite over support caps $S\in\{1,2,3,4,8\}$.

\section{Numerical Results}\label{sec:results}
Across $\PaperObservationCount{}$ single-snapshot evaluations spanning $\PaperConditionCount{}$ cells, all receivers run with strictly frozen parameters; positive relative gains denote percentage reductions in 3-D RMSE.

\subsection{Core Positioning Accuracy, Waveform Physics, and Latency}
\begin{figure*}[!t]
\centering
\subfloat[Fixed state]{\includegraphics[width=.455\textwidth]{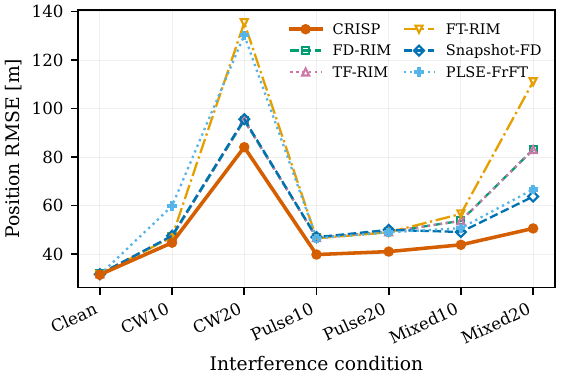}}\hspace{0.03\textwidth}%
\subfloat[Displaced state]{\includegraphics[width=.455\textwidth]{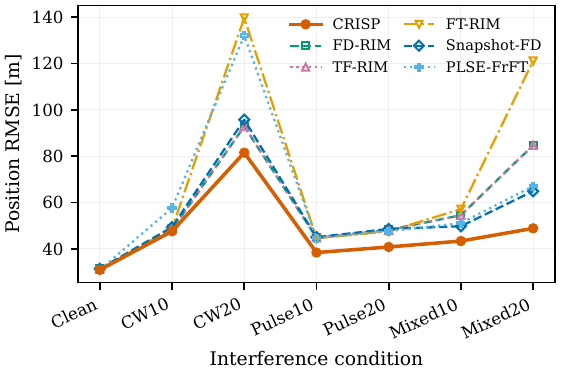}}
\caption{Positioning RMSE across individual core interference environments in fixed and displaced states.}
\label{fig:core}
\end{figure*}

Table~\ref{tab:main} reports accuracy, tail metrics, and serial runtime across the $\CoreTrials{}$-trial core benchmark split evenly between $\FixedTrials{}$ fixed-state and $\DisplacedTrials{}$ displaced-state trials, while Fig.~\ref{fig:core} resolves RMSE by interference type. \method{} attains the lowest error on every metric, lowering pooled RMSE from \CoreSnapshotRMSE{} m under Snapshot-FD to \CoreCRISPRMSE{} m for a \CoreRMSEGain{}\% reduction with simultaneous $95\%$ CI $[\CoreGainCILow{},\CoreGainCIHigh{}]\%$, alongside gains of \FixedRMSEGain{}\% from \FixedSnapshotRMSE{} to \FixedCRISPRMSE{} m in the fixed state and \DisplacedRMSEGain{}\% from \DisplacedSnapshotRMSE{} to \DisplacedCRISPRMSE{} m in the displaced state. Tail error P99 drops by more than $30$ m relative to Snapshot-FD, and gross-error probability $P(e>100\,\mathrm{m})$ falls from \CoreSnapshotTailRate{}\% to \CoreCRISPTailRate{}\%, a \CoreTailReduction{}\% reduction shown in Fig.~\ref{fig:tailcomparison}(a). Receiver clock-bias RMSE likewise decreases by $14.66\%$ from \CoreSnapshotClockRMSE{} m, or $119.34$ ns, under Snapshot-FD to \CoreCRISPClockRMSE{} m, or $101.85$ ns, compared with $38.06$ m for FD/TF-RIM, $42.37$ m for PLSE-FrFT, and $47.46$ m for FT-RIM, while $\leq100$-m capture rises from $93.25\%$ to $96.04\%$ and final-estimate search-boundary hits equal $0.00\%$ versus $0.04\%$ to $0.82\%$ across single-branch baselines.

\begin{table*}[!t]
\centering\footnotesize
\caption{Core Benchmark Positioning Accuracy and Serial Processing Cost}
\label{tab:main}
\setlength{\tabcolsep}{3.5pt}
\begin{tabular*}{\textwidth}{@{\extracolsep{\fill}}lrrrrrrrrrr@{}}
\toprule
Method & \shortstack{Fixed\\(m)} & \shortstack{Displaced\\(m)} & \shortstack{Pooled\\(m)} & \shortstack{P95\\(m)} & \shortstack{P99\\(m)} & $P(e>100\,\mathrm{m})$ & \shortstack{Mean\\solves} & \shortstack{Mean\\(ms)} & \shortstack{Median\\(ms)} & \shortstack{P95\\(ms)} \\
\midrule
CRISP & \best{50.40} & \best{49.67} & \best{50.03} & \best{94.43} & \best{136.89} & \best{3.96\%} & 1.43 & 82.63 & 52.21 & 275.56 \\
Snapshot-FD & \second{58.01} & \second{58.13} & \second{58.07} & \second{108.30} & \second{167.41} & \second{6.75\%} & 1.00 & \second{44.01} & 41.16 & 66.65 \\
TF-RIM & 61.72 & 61.39 & 61.55 & 118.11 & 174.95 & 8.24\% & 1.00 & 44.65 & \second{41.00} & \best{61.67} \\
FD-RIM & 61.72 & 61.40 & 61.56 & 118.11 & 174.95 & 8.26\% & 1.00 & \best{43.94} & \best{40.61} & \second{63.11} \\
PLSE-FrFT & 68.76 & 68.74 & 68.75 & 133.41 & 227.60 & 11.24\% & 1.00 & 127.25 & 124.19 & 223.94 \\
FT-RIM & 77.11 & 80.21 & 78.68 & 153.65 & 281.29 & 12.65\% & 1.00 & 47.16 & 43.19 & 79.83 \\
\bottomrule
\end{tabular*}
\end{table*}

Figure~\ref{fig:core} illustrates the underlying waveform mechanisms. Under Clean conditions, \method{} preserves unclipped spread-spectrum correlation energy. Under Pulse10, Pulse20, Mixed10, and Mixed20, frequency-domain FD-RIM and Snapshot-FD spread transient burst energy across DFT bins within the $2$-ms or $0.25$-ms transform window, raising spectral noise floors, whereas cascaded FT-RIM over-clips clean chips under $2$-bit scale collapse and PLSE-FrFT suffers ridge-model mismatch on broadband pulses. Descriptor routing dispatches pulsed and compound snapshots to localized I/Q-STFT blanking or matched cascades that excise bursts while preserving uncorrupted chips. Under CW20, $2$-bit AGC compression generates odd-order quantization harmonics that leak past fixed spectral clippers, whereas Newton-refined coherent line projection and multibranch fusion suppress sidelobe-induced $100$-m DPE outliers.

\begin{figure*}[!t]
\centering
\subfloat[Accuracy--latency frontier]{\includegraphics[width=.455\textwidth]{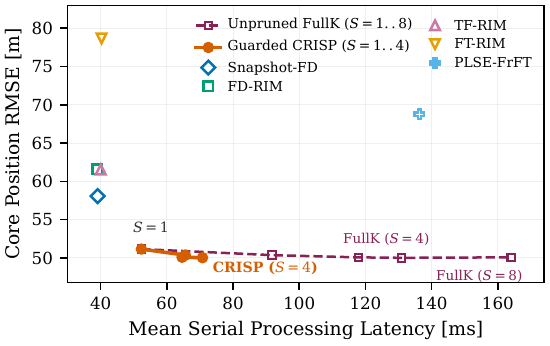}}\hspace{0.03\textwidth}%
\subfloat[Stratified latency breakdown]{\includegraphics[width=.455\textwidth]{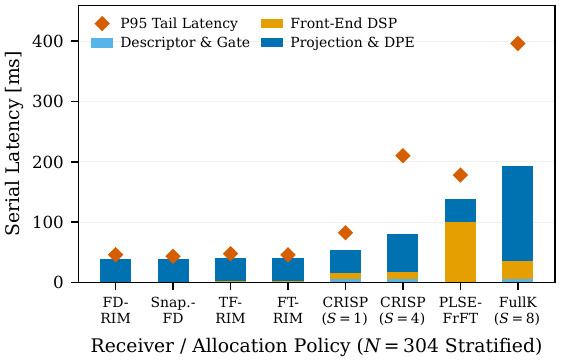}}
\caption{Accuracy--latency Pareto frontier across support caps $S\in\{1,2,3,4,8\}$ on the core benchmark and per-stage serial runtime decomposition across the $304$-snapshot stratified suite.}
\label{fig:latencybudget}
\end{figure*}

Runtime measurements in Table~\ref{tab:main} and across the $304$-snapshot stratified suite of $224$ core and $80$ high-jamming stress snapshots in Fig.~\ref{fig:latencybudget} quantify the accuracy-latency trade-off of selective fusion. \method{} averages \CoreMeanSolves{} DPE solves on the core benchmark and \OverallMeanSolves{} solves across all $\PaperObservationCount{}$ test snapshots, invoking multibranch fusion on $17.03\%$ of trials. On the $56$-snapshot stream in Table~\ref{tab:main}, median runtime is $52.21$ ms versus $41.16$ ms for Snapshot-FD, while four-branch activation in Bin~2 raises mean latency to \CRISPMeanRuntime{} ms, or $1.88\times$ Snapshot-FD at \SnapshotMeanRuntime{} ms and $35\%$ below PLSE-FrFT at $127.25$ ms, with P95 latency reaching $275.56$ ms, or $4.13\times$ Snapshot-FD at $66.65$ ms. Across the $304$-snapshot suite, \method{} averages $80.24$ ms with median $47.04$ ms and P95 $210.08$ ms at $1.63$ solves, spanning $70.77$ ms on the core stratum and $106.77$ ms at $2.20$ solves on the stress stratum, compared with $39.17$ ms mean and $43.32$ ms P95 for Snapshot-FD and $138.42$ ms mean for PLSE-FrFT. Stage decomposition in Fig.~\ref{fig:latencybudget}(b) attributes $5.16$ ms to descriptor extraction and gating, $11.50$ ms to active front-end DSP versus $1.28$ ms for Snapshot-FD and $100.09$ ms for PLSE-FrFT, and $63.56$ ms to projection, DPE, and fusion. Tracing support caps $S\in\{1,2,3,4\}$ in Fig.~\ref{fig:latencybudget}(a) lowers guarded core RMSE from $51.16$ m at $S=1$ and $52.30$ ms core mean to $50.37$ m at $S=2$ and $65.58$ ms, $50.09$ m at $S=3$ and $64.60$ ms, and \CoreCRISPRMSE{} m at $S=4$ and $70.77$ ms, plotted alongside the unpruned \textit{FullK} trajectory across $S\in\{1,2,3,4,8\}$. At $S=8$, guarded \method{} reaches $50.09$ m with $1.86$ core solves, whereas unpruned \textit{FullK} reaches $50.08$ m with $3.29$ core solves, $164.04$ ms core mean latency, and $396.06$ ms pooled P95 latency, consistent with finite-sample simplex variability and identifying $S=4$ as the accuracy-latency knee alongside low-latency operating points $S\in\{1,2\}$.

\subsection{Constellation Geometry Transfer and Anisotropic Fusion}
\begin{figure*}[!t]\centering
\subfloat[Fixed and displaced states]{\includegraphics[width=.455\textwidth]{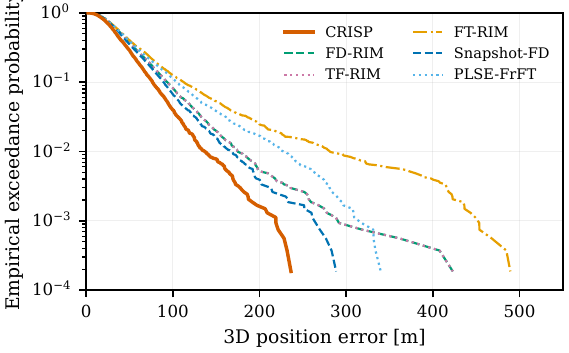}}\hspace{0.03\textwidth}%
\subfloat[Geometry transfer]{\includegraphics[width=.455\textwidth]{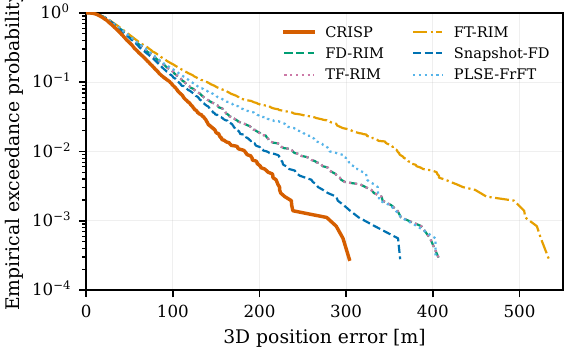}}
\caption{Empirical complementary cumulative distributions of three-dimensional positioning error in the core benchmark and across changed geometries.}\label{fig:tailcomparison}
\end{figure*}

Table~\ref{tab:geometry} and Fig.~\ref{fig:geometrycombined}(b) evaluate zero-retraining transfer of the G0-calibrated receiver across constellations G1 through G4 at $896$ snapshots per geometry, with tail curves in Fig.~\ref{fig:tailcomparison}(b). Across the $\GeometryTrials{}$-snapshot G1--G4 pool, \method{} lowers RMSE from \GeometrySnapshotRMSE{} m under Snapshot-FD to \GeometryCRISPRMSE{} m, a \GeometryRMSEGain{}\% reduction with simultaneous $95\%$ CI $[\GeometryGainCILow{},\GeometryGainCIHigh{}]\%$, achieving per-geometry reductions of $11.48\%$, $13.23\%$, $11.16\%$, and $14.98\%$ on G1 through G4 while cutting $P(e>100\,\mathrm{m})$ by \GeometryTailReduction{}\% from \GeometrySnapshotTailRate{}\% to \GeometryCRISPTailRate{}\%. On the $\HoldoutTrials{}$ unseen G5 trials at PDOP $2.15$, \method{} reduces RMSE by \HoldoutRMSEGain{}\% from \HoldoutSnapshotRMSE{} to \HoldoutCRISPRMSE{} m with $95\%$ CI $[\HoldoutGainCILow{},\HoldoutGainCIHigh{}]\%$ and cuts $>100$-m errors by \HoldoutTailReduction{}\%.

\begin{table}[!t]
\centering\footnotesize
\caption{Positioning RMSE in Meters Across Changed Satellite Geometries}
\label{tab:geometry}
\setlength{\tabcolsep}{3.6pt}
\begin{tabular*}{\columnwidth}{@{\extracolsep{\fill}}lrrrrr@{}}
\toprule
Method & G1 & G2 & G3 & G4 & Pool \\
\midrule
PDOP & 2.26 & 2.38 & 4.33 & 2.43 & -- \\
\midrule
CRISP & \best{46.90} & \best{50.19} & \best{88.11} & \best{50.08} & \best{61.22} \\
Snapshot-FD & \second{52.98} & \second{57.85} & \second{99.17} & \second{58.90} & \second{69.75} \\
TF-RIM & 55.45 & 62.02 & 106.65 & 66.90 & 75.45 \\
FD-RIM & 55.47 & 62.03 & 106.65 & 66.92 & 75.46 \\
PLSE-FrFT & 65.56 & 72.08 & 112.35 & 69.96 & 82.17 \\
FT-RIM & 79.65 & 83.10 & 130.11 & 84.36 & 96.56 \\
\bottomrule
\end{tabular*}
\end{table}

Single-sided layout G3 at PDOP $4.33$ separates operator-level cross-risk cancellation from coordinate weight anisotropy. Clustering all six satellites within $20^\circ$ to $155^\circ$ azimuth degrades cross-sector observability, where \method{} lowers RMSE from $99.17$ m under Snapshot-FD and $106.65$ m under FD-RIM to $88.11$ m. Matched scalar versus coordinate calibration in Table~\ref{tab:matched} pinpoints each mechanism's contribution: on calibration geometry G0, coordinate-specific \textit{Core} at $50.03$ m improves upon isotropic \textit{ScalarGuarded} at $50.31$ m by $0.55\%$ with $95\%$ CI $[0.22,0.87]\%$, reaching a $1.38\%$ gain from $83.96$ to $82.80$ m with $95\%$ CI $[0.57,2.17]\%$ in Bin~2 by assigning $17.1\%$ auxiliary weight to the vertical Up axis versus $2.7\%$ on East and $7.7\%$ on North. Under zero-retraining transfer to G3, over $90\%$ of the $11.06$-m gain over Snapshot-FD stems from descriptor-conditioned operator selection at $89.14$ m and shared-support correlation cancellation in \textit{ScalarGuarded} at $87.18$ m: frozen G0 weights still lower G3 North error from $15.12$ to $14.97$ m and Up error from $43.10$ to $42.75$ m, whereas G3 concentrates horizontal ill-conditioning along East at $75.58$ versus $74.26$ m, leaving \textit{ScalarGuarded} $0.93$ m ahead of \textit{Core} on G3 and $0.25$ m ahead across the G1--G4 pool at $60.96$ versus $61.22$ m. Coordinate anisotropy thus improves accuracy when directional DOP structure aligns with calibration such as persistent vertical ill-conditioning, while scalar cross-risk weights transfer more uniformly across uncalibrated horizontal sector shifts.

\begin{figure*}[t]
\centering
\CRISPsetrowheight{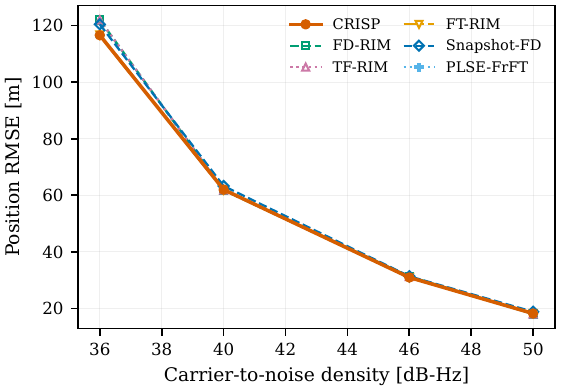}%
\subfloat[Clean]{\includegraphics[height=\CRISProwheight]{cn0_Clean.pdf}}\hfill
\subfloat[CW20]{\includegraphics[height=\CRISProwheight]{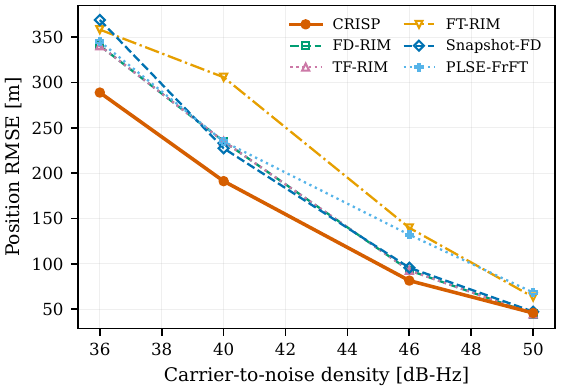}}\hfill
\subfloat[Mixed20]{\includegraphics[height=\CRISProwheight]{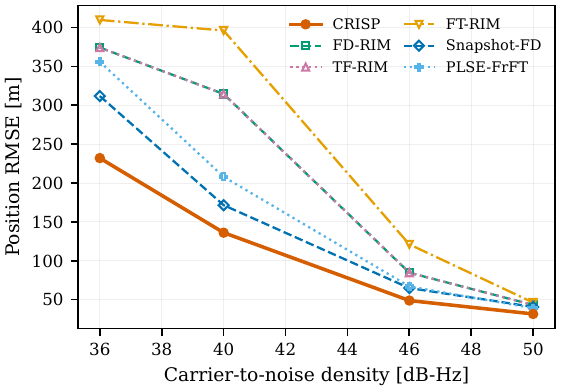}}
\caption{Positioning RMSE sensitivity across carrier-to-noise density ratios $C/N_0$.}
\label{fig:cn0}
\end{figure*}

\begin{figure*}[t]
\centering
\CRISPsetrowheight{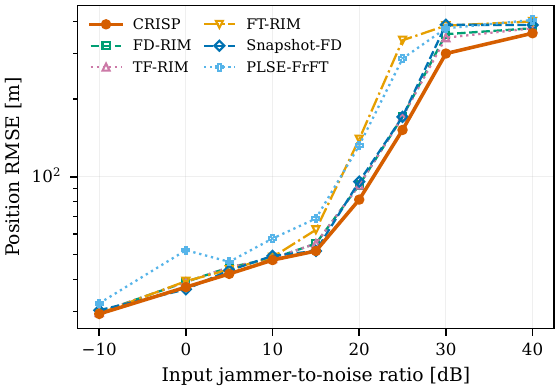}%
\subfloat[CW]{\includegraphics[height=\CRISProwheight]{jnr_CW.pdf}}\hfill
\subfloat[Pulse]{\includegraphics[height=\CRISProwheight]{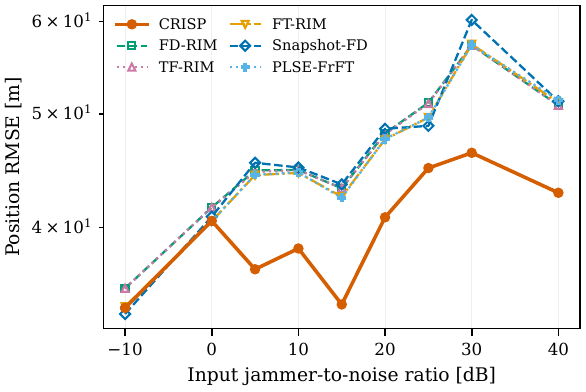}}\hfill
\subfloat[Mixed]{\includegraphics[height=\CRISProwheight]{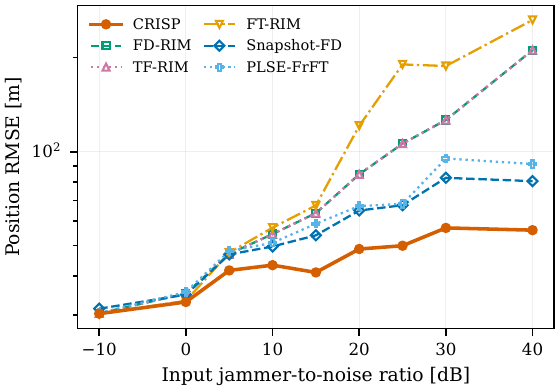}}
\caption{Positioning RMSE across input jammer-to-noise ratios with $2$-bit conversion and $C/N_0=46$ dB-Hz.}
\label{fig:jnr}
\end{figure*}

\begin{figure*}[t]
\centering
\CRISPsetrowheight{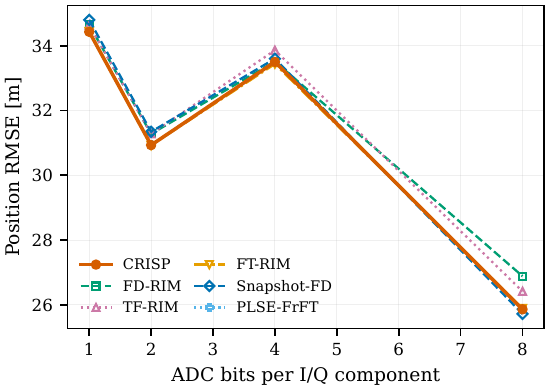}%
\subfloat[Clean]{\includegraphics[height=\CRISProwheight]{adc_Clean.pdf}}\hfill
\subfloat[CW20]{\includegraphics[height=\CRISProwheight]{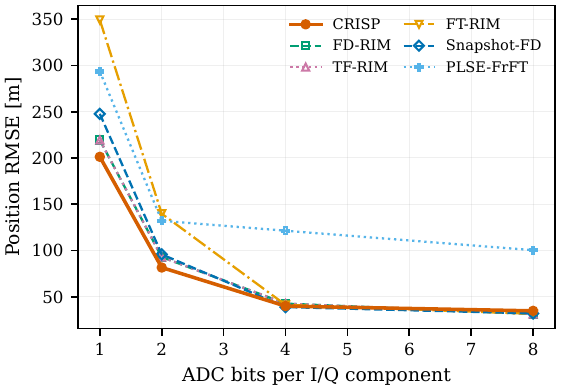}}\hfill
\subfloat[Mixed20]{\includegraphics[height=\CRISProwheight]{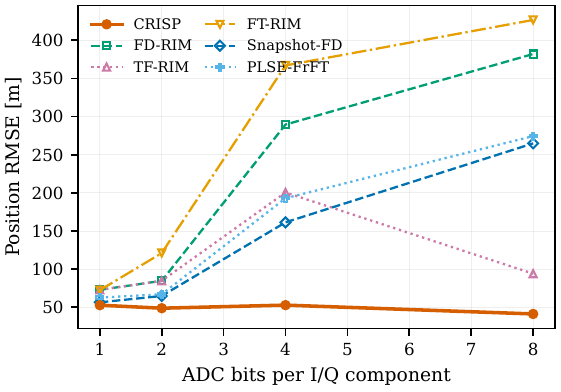}}
\caption{Zero-retraining transfer across ADC bit resolutions from $1$ to $8$ bits per I/Q component.}
\label{fig:adc}
\end{figure*}

\begin{figure*}[t]
\centering
\CRISPsetrowheight{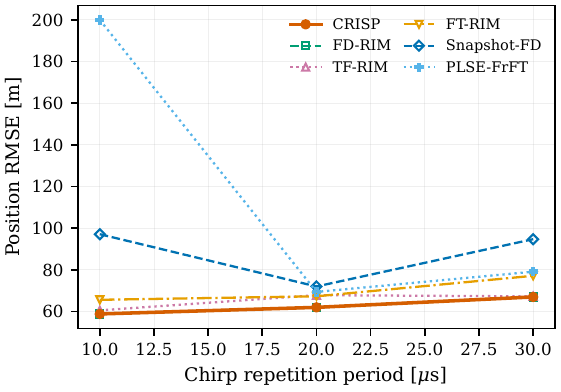}%
\subfloat[LFM, $B=2$, J/N $=10$ dB]{\includegraphics[height=\CRISProwheight]{lfm_B2_J10.pdf}}\hfill
\subfloat[QFM, $B=16$, J/N $=10$ dB]{\includegraphics[height=\CRISProwheight]{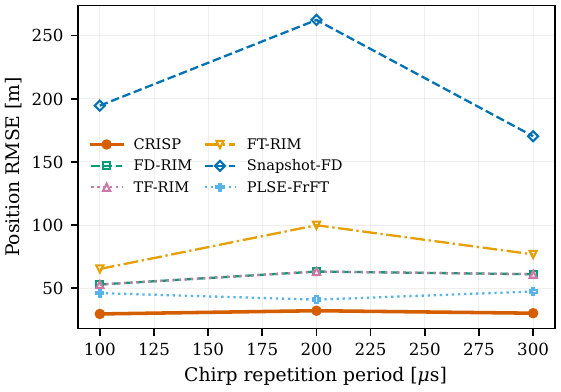}}\hfill
\subfloat[Frequency hopping]{\includegraphics[height=\CRISProwheight]{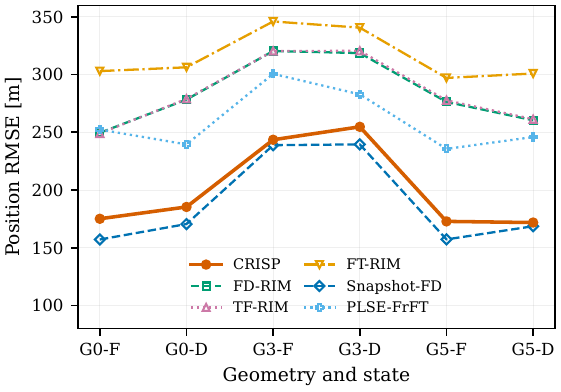}}
\caption{Frozen-receiver stress evaluations across LFM/QFM chirp repetition periods and unseen fast frequency hopping.}
\label{fig:pressure}
\end{figure*}

\subsection{Sparse Allocation, Validation Screening, and Cross-Moment Ablation}
Across the $\FitTrialCount{}$ risk-fit and $\GuardTrialCount{}$ validation snapshots partitioned into $M=7$ pre-frozen bins, criterion \eqref{eq:calguard} prunes six bins to single matched front ends: bin~0 with $267$ fit and $119$ guard trials selects Branch~1 at score $+2.14\text{ m}^2$; bin~1 with $255$ and $126$ trials selects Branch~8 at $+13.08\text{ m}^2$; bin~4 with $212$ and $114$ trials selects Branch~5 at $0.00\text{ m}^2$; and bins~3, 5, and~6 with $308/150$, $251/144$, and $240/133$ trials select Branch~6 at $+0.46$, $+2.68$, and $+0.08\text{ m}^2$, yielding the \CoreMeanSolves{}-solve core average. Bin~2 with $259$ fit and $110$ guard trials captures severe compound AGC-quantizer harmonic distortion where residual errors across Branches~1, 2, 4, and~8 compensate: on its $110$ validation trials, $\overline D=-924.11\text{ m}^2$ with $\widehat{\operatorname{SE}}(\overline D)=344.94\text{ m}^2$, guard score $-356.69\text{ m}^2$, median $-63.38\text{ m}^2$, and one-sided Wilcoxon $p=0.062$, whereas the valid bounded empirical Bernstein upper bound \cite{maurer2009empirical} over $[-300,300]^3$ m remains conservatively positive at $+3.07\times 10^4\text{ m}^2$ for $n_{\mathrm{val}}=110$. Across the $\BinTwoTrials{}$ core test snapshots in bin~2, this four-branch policy lowers RMSE from \BinTwoSingleRMSE{} m to \BinTwoCoreRMSE{} m, a \BinTwoGain{}\% reduction with simultaneous $95\%$ CI $[\BinTwoGainCILow{},\BinTwoGainCIHigh{}]\%$. Continuous $10$-D feature routers \textit{RidgeRouter-Single} and \textit{kNNRouter-Single} yield the identical $87.47$-m RMSE in bin~2 and $51.18$ and $51.16$ m across Core because both fitted routers select Branch~8 across all $\BinTwoTrials{}$ bin~2 test snapshots, leaving four-branch fusion $5.34\%$ ahead with marginal $95\%$ CI $[3.62,6.99]\%$. Refitting $(\widehat{\boldsymbol\mu},\widehat{\boldsymbol S},\{\widehat{\boldsymbol c}_m\}_{m=0}^6)$ from scratch on $\RepFitTrialCount{}$ disjoint label-free snapshots in Table~\ref{tab:provenance} reproduces the identical six-single, one-quadruple structure at $1.43$ core solves and yields $50.03$ m on Core, $61.21$ m across G1--G4, and $45.06$ m on G5.

Table~\ref{tab:matched} contrasts matched-library policies across Core, G3, and G5 while distinguishing validation-pruned \textit{Core} from unpruned \textit{FullK}. Moving from \textit{Single} at \MatchedSingleRMSE{} m and $1.00$ solve to unpruned \textit{FullK} at \MatchedFullKRMSE{} m and $2.72$ solves reduces Core RMSE by \MatchedFusionGain{}\% with simultaneous $95\%$ CI $[1.31,3.16]\%$, while \textit{Core} achieves $50.03$ m, a $2.19\%$ gain over \textit{Single}, at $47\%$ fewer solves. Omitting off-diagonal correlations in \textit{Diagonal} at $4.00$ solves degrades Core RMSE to $53.05$ m, confirming \eqref{eq:direction}. \textit{CenteredCov} at $3.00$ solves reaches $50.12$ m on Core and $45.29$ m on G5, with simultaneous $95\%$ CIs $[-0.01,0.40]\%$ and $[-0.06,1.00]\%$ relative to \textit{FullK} spanning zero under symmetric zero-mean state perturbations. Uncentered moments retain exact algebraic alignment with uncorrected position MSE when directional biases arise: in a controlled G3 diagnostic with coherent $0.35$-chip in-phase reflections calibrated on $288$ snapshots and evaluated on $384$ disjoint test snapshots inducing $46.8$ to $71.3$ m branch biases, \textit{FullK} reaches $95.51$ m versus $96.49$ m for \textit{CenteredCov} and $97.57$ m for \textit{Single}, corresponding to a $1.02\%$ reduction over \textit{CenteredCov} with marginal $95\%$ CI $[-0.35,2.34]\%$.

\begin{table}[!t]
\centering\footnotesize
\caption{Matched-Library Allocation, Weighting, and Moment Ablation in Meters}
\label{tab:matched}
\setlength{\tabcolsep}{2.3pt}
\begin{tabular*}{\columnwidth}{@{\extracolsep{\fill}}lrrrl@{}}
\toprule
\multicolumn{5}{c}{Core Benchmark, $n=\MatchedTrialCount{}$} \\
\midrule
Policy & Solves & RMSE & FullK Gain & 95\% CI \\
\midrule
Single & 1.00 & 51.16 & 2.23\% & [1.31, 3.16]$^\dagger$ \\
RidgeRouter-Single & 1.00 & 51.18 & 2.26\% & [1.54, 2.98] \\
kNNRouter-Single & 1.00 & 51.16 & 2.24\% & [1.52, 2.96] \\
Diagonal & 4.00 & 53.05 & 5.72\% & [4.38, 7.06]$^\dagger$ \\
ScalarGuarded & 1.43 & 50.31 & 0.58\% & [0.25, 0.90] \\
ScalarFullK & 1.86 & 50.30 & 0.55\% & [0.23, 0.87] \\
CenteredCov & 3.00 & 50.12 & 0.19\% & [$-0.01$, 0.40]$^\dagger$ \\
Core & 1.43 & \second{50.03} & 0.03\% & [$-0.00$, 0.07] \\
FullK & 2.72 & \best{50.02} & -- & -- \\
\midrule
\multicolumn{5}{c}{Single-Sided Geometry G3, $n=896$, PDOP $4.33$} \\
\midrule
Policy & Solves & RMSE & FullK Gain & 95\% CI \\
\midrule
Single & 1.00 & 89.14 & 1.16\% & [0.38, 1.93]$^\dagger$ \\
Diagonal & 4.00 & 92.22 & 4.47\% & [1.88, 6.83]$^\dagger$ \\
ScalarGuarded & 1.43 & \second{87.18} & $-1.06\%$ & [$-1.71$, $-0.43$] \\
ScalarFullK & 1.86 & \best{87.14} & $-1.10\%$ & [$-1.75$, $-0.48$] \\
CenteredCov & 3.00 & 88.13 & 0.03\% & [$-0.23$, 0.29]$^\dagger$ \\
Core & 1.43 & 88.11 & 0.01\% & [$-0.19$, 0.20] \\
FullK & 2.73 & 88.10 & -- & -- \\
\midrule
\multicolumn{5}{c}{G5 Holdout, $n=\HoldoutTrials{}$, PDOP $2.15$} \\
\midrule
Policy & Solves & RMSE & FullK Gain & 95\% CI \\
\midrule
Single & 1.00 & 45.03 & $-0.07\%$ & [$-2.36$, 2.25]$^\dagger$ \\
Diagonal & 4.00 & 46.89 & 3.89\% & [$-0.23$, 8.20]$^\dagger$ \\
ScalarGuarded & 1.43 & \second{44.83} & $-0.53\%$ & [$-1.22$, 0.16] \\
ScalarFullK & 1.86 & \best{44.82} & $-0.55\%$ & [$-1.23$, 0.13] \\
CenteredCov & 3.00 & 45.29 & 0.49\% & [$-0.06$, 1.00]$^\dagger$ \\
Core & 1.43 & 45.06 & $-0.01\%$ & [$-0.13$, 0.10] \\
FullK & 2.73 & 45.06 & -- & -- \\
\bottomrule
\multicolumn{5}{@{}>{\raggedright\arraybackslash}p{\columnwidth}@{}}{{\scriptsize $^\dagger$Simultaneous $95\%$ Bonferroni CI; other rows show paired marginal $95\%$ CIs.}}
\end{tabular*}
\end{table}

\subsection{Online Reliability Gating, Parametric Sweeps, and Stress Limits}
Nominal core and G1--G5 trials satisfy $d_c^2\leq\tau_d$ and retain the core policy, whereas parametric stress suites trigger $\RerouteCount{}$ online reroutes across $\PaperObservationCount{}$ observations. Table~\ref{tab:rg} reports paired same-snapshot outcomes via $\Delta\operatorname{RMSE}=\operatorname{RMSE}_{\mathrm{off}}-\operatorname{RMSE}_{\mathrm{on}}$, where RG lowers pooled RMSE by \RerouteReduction{} m from \RerouteOffRMSE{} to \RerouteOnRMSE{} m with $95\%$ CI $[28.43,38.35]$ m. Quantizer-occupancy and proper burst-weighting routes provide strong protection under extreme J/N, $1$-bit quantization, and rapid chirps, cutting RMSE by $135.40$ m under frequency Huber and by $22.00$, $17.46$, and $14.06$ m under $256$-point STFT, $64$-point STFT, and local-power weighting. Consistent with \eqref{eq:gaterisk}, net gate performance reflects the evaluated regime mix, in which $2,390$ protective reroutes outweigh $716$ adverse low-lag reroutes into Identity and four-line projection.

\begin{table}[!t]
\centering\footnotesize
\caption{Paired Same-Snapshot Reliability Gate Performance in Meters}
\label{tab:rg}
\setlength{\tabcolsep}{2.1pt}
\begin{tabular*}{\columnwidth}{@{\extracolsep{\fill}}lrrrl@{}}
\toprule
Routing action & $n$ & RG off & RG on & \shortstack{$\Delta$RMSE\\{[95\% CI]}} \\
\midrule
Core retention & 14\,462 & \best{90.77} & \best{90.77} & \phantom{$-$}0.00 [0.00, 0.00] \\
\midrule
Local-power weighting & 256 & \second{60.99} & \best{46.93} & \phantom{0}14.06 [11.77, 16.40] \\
Identity & 394 & \best{88.70} & \second{134.12} & $-45.42$ [$-59.11$, $-32.57$] \\
Four-line projection & 322 & \best{77.51} & \second{148.55} & $-71.04$ [$-85.70$, $-56.69$] \\
\midrule
256-point STFT & 256 & \second{74.77} & \best{52.76} & \phantom{0}22.00 [17.27, 26.84] \\
Frequency Huber & 640 & \second{242.20} & \best{106.80} & 135.40 [123.63, 147.42] \\
64-point STFT & 1\,238 & \second{149.20} & \best{131.74} & \phantom{0}17.46 [12.06, 22.88] \\
\midrule
All actual reroutes & 3\,106 & \second{152.80} & \best{119.37} & \phantom{0}33.43 [28.43, 38.35] \\
\bottomrule
\end{tabular*}
\end{table}

Figures~\ref{fig:cn0}--\ref{fig:pressure} trace parametric sensitivity across $C/N_0$, J/N up to $40$ dB, $1$- to $8$-bit quantization, and LFM/QFM chirps, where \method{} tracks the lower error envelope across almost all cells. Under CW20 at $B=8$ bits in Fig.~\ref{fig:adc}(b) and at $C/N_0=50$ dB-Hz in Fig.~\ref{fig:cn0}(b), FD-RIM edges ahead by $3.35$ m at $31.33$ versus $34.68$ m and by $1.09$ m at $44.78$ versus $45.87$ m because descriptor centroids calibrated under $2$-bit AGC clipping harmonics shift slightly under linear $8$-bit conversion. The adverse Identity and four-line rows in Table~\ref{tab:rg} and the $\HoppingTrials{}$-trial fast frequency-hopping test in Fig.~\ref{fig:pressure}(c) delineate an explicit operating boundary of the low-lag gate: phase-continuous $20$ to $100\,\mu\text{s}$ dwells step across $20$ to $100$ frequencies per $2$-ms window, equalizing quarter-block power to $r\leq\tau_r$ while decorrelating adjacent samples to $c_1<\tau_c$, which triggers Identity or four-line projection designed for weak broadband shifts or sparse tones and raises RMSE by $44.17$ m from $159.41$ m under ungated \textit{Core} to \HoppingCRISPRMSE{} m with $95\%$ CI $[35.99,52.59]$ m versus \HoppingSnapshotRMSE{} m for Snapshot-FD. Under unmodeled $0.2$-, $0.5$-, and $1.0$-chip specular multipath, \method{} attains $73.12$, $114.64$, and $53.67$ m versus $75.04$, $115.36$, and $53.17$ m for Snapshot-FD, reflecting code-phase bias inside the replica model.

\begin{figure*}[!t]
\centering
\CRISPsetrowheight{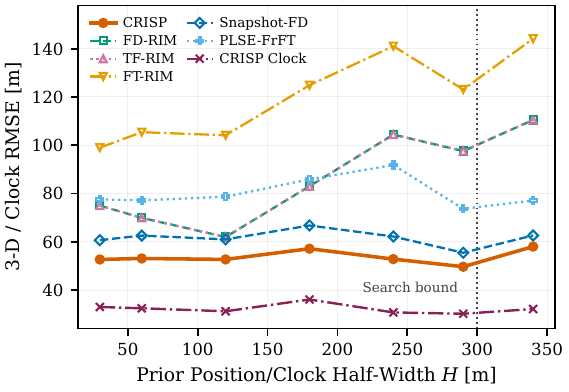}%
\subfloat[Prior uncertainty sweep]{\includegraphics[height=\CRISProwheight]{aiding_prior_rmse.pdf}}\hfill
\subfloat[Capture and boundary hits]{\includegraphics[height=\CRISProwheight]{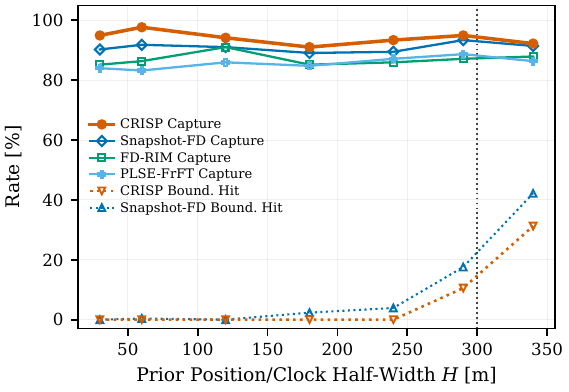}}\hfill
\subfloat[Doppler error sweep]{\includegraphics[height=\CRISProwheight]{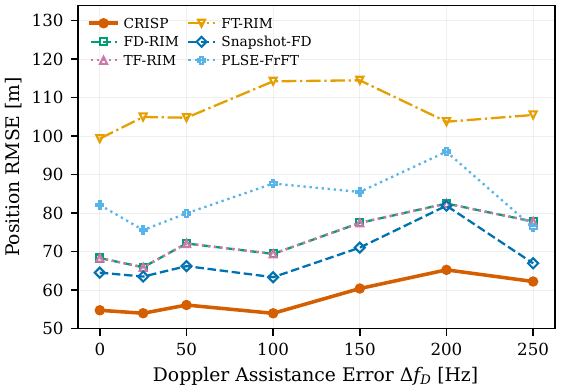}}
\caption{Assisted operating-range sensitivity across Clean, CW20, Pulse20, and Mixed20 snapshots at $256$ paired trials per point over the $[-300,300]^4$-m search box, showing position and clock RMSE, $\leq100$-m capture and boundary hit rates across truth half-widths $H$, and position RMSE across Doppler errors $\Delta f_D$.}
\label{fig:aidingsweep}
\end{figure*}

\subsection{Spatial-Transfer Evidence and Assisted Operating Range}
Static location-prior diagnostics verify waveform-driven positioning across displaced states. Returning the reference coordinate $[23,-31,14]^{\mathsf T}$ m yields $0.00$ m error on fixed trials and $59.84$ m on $\pm60$-m displaced trials for a degenerate $42.32$-m pooled value, or $42.35$ m for the calibration mean, whereas \method{} reaches \DisplacedCRISPRMSE{} m on displaced trials. Expanding truth half-widths to $120$ and $180$ m yields $61.68$ and $56.03$ m for \method{} while static-prior errors surge to $120.04$ and $170.07$ m. Across $1,152$ translated-center trials at three shifted zones, \method{} sustains $56.53$, $62.07$, and $55.80$ m for a $58.20$-m pooled RMSE versus $162.08$ m for the calibration-mean prior, demonstrating empirical spatial transfer when $\Theta$ centers on the coarse prior alongside the fixed-weight affine equivariance in Section~\ref{sec:gatinggeom}.

Figure~\ref{fig:aidingsweep} quantifies the assisted operating range across a standalone $4,608$-snapshot suite spanning $18$ sweep points of $256$ paired trials each across Clean, CW20, Pulse20, and Mixed20 on G0. In Fig.~\ref{fig:aidingsweep}(a) and Fig.~\ref{fig:aidingsweep}(b), uniform 4-D prior half-widths $H\in\{30,60,120,180,240\}$ m inside $[-300,300]^4$ m yield $52.67$ to $57.13$ m position RMSE, $30.72$ to $36.08$ m clock RMSE, $91.02\%$ to $97.66\%$ capture within $100$ m, and $0.00\%$ final-estimate boundary hits for \method{}, outperforming Snapshot-FD at $60.66$ to $66.74$ m position RMSE and up to $3.91\%$ boundary hits while simplex fusion suppresses constituent-branch boundary contacts occurring on $12.50\%$ to $20.31\%$ of pooled trials. At $H=290$ and $340$ m near or beyond $\pm300$ m, final boundary clamping rises to $10.55\%$ and $31.25\%$ for \method{} and $17.58\%$ and $42.19\%$ for Snapshot-FD, establishing sub-$250$-m position-clock assistance as the operational domain of the $600$-m search box. Across uncompensated uniform carrier-Doppler errors $\Delta f_D\in\{0,25,50,100,150,200,250\}$ Hz in Fig.~\ref{fig:aidingsweep}(c) up to half the $500$-Hz coherent first-null spacing, \method{} moves smoothly from $54.77$ m position RMSE and $33.63$ m clock RMSE at $0$ Hz to $53.98$ m at $100$ Hz, $60.42$ m at $150$ Hz, $65.27$ m at $200$ Hz, and $62.24$ m at $250$ Hz, where $250$-Hz $\leq100$-m capture remains $89.84\%$ with $0.00\%$ final boundary hits, consistently ahead of Snapshot-FD at $63.36$ to $81.93$ m and PLSE-FrFT at $75.61$ to $95.97$ m. Zero-mean Gaussian ephemeris range biases $\sigma_{\mathrm{eph}}\in\{0,2,5,10\}$ m similarly yield $51.78$ to $58.43$ m position RMSE, $32.26$ to $34.91$ m clock RMSE, and $0.00\%$ boundary hits versus $61.09$ to $65.61$ m for Snapshot-FD \cite{vandiggelen2009agps}.

\FloatBarrier
\section{Conclusion}\label{sec:conclusion}
This paper presented \method{}, an aided single-snapshot GNSS positioning architecture coupling nonlinear interference mitigation with direct position estimation via sparse cross-risk calibration. Coordinate-simplex weighting over uncentered second-order error moments on a pre-frozen waveform partition, combined with validation screening and same-snapshot reliability gating, retains four-branch fusion under severe AGC-quantizer distortion while defaulting to single matched operators elsewhere. Theoretical analysis established the post-suppression DPE contrast decomposition, a conditional finite-sample excess-risk bound across disjoint calibration splits, and an exact gating risk identity. Across $\PaperObservationCount{}$ simulated $2$-ms GPS L1 C/A snapshots, \method{} reduced core RMSE by \CoreRMSEGain{}\%, $>100$-m errors by \CoreTailReduction{}\%, and multi-geometry RMSE by \GeometryRMSEGain{}\% over Snapshot-FD at \CoreMeanSolves{} core and \OverallMeanSolves{} overall solves. Future work prioritizes testing frozen-calibration \method{} policies on RF and intermediate-frequency replay with interference injected before physical AGC and low-bit analog-to-digital conversion, distinguishing hardware front-end validation from semi-synthetic tests that digitally add interference and requantize high-resolution recordings, alongside hopping-aware descriptors and multipath replica models.

\appendices
\section{Single-Snapshot Descriptor and Gate}\label{app:descriptor}
From the $N$-sample quantized block $\boldsymbol y$ shared by all processing paths, define normalized instantaneous power and DFT periodogram ordinates using a numerical floor of $10^{-12}$:
\begin{equation}
 v_n=\frac{|y_n|^2}{N^{-1}\sum_m|y_m|^2},\qquad
 P_k=\frac{|[\boldsymbol F\boldsymbol y]_k|^2}{N^{-1}\sum_l|[\boldsymbol F\boldsymbol y]_l|^2}.
\end{equation}
Let $q_a(P)$ denote the empirical $a$-quantile of $\{P_k\}$, $s_r$ the mean of $v_n$ over temporal quarter $r\in\{1,2,3,4\}$, and $c_h$ the normalized lag-$h$ autocorrelation magnitude
\begin{equation}
 c_h=\frac{\left|(N-h)^{-1}\sum_{n=h}^{N-1}y_n y_{n-h}^*\right|}
                    {N^{-1}\sum_n|y_n|^2}.
\end{equation}
The ten stored descriptor components are ordered as
\begin{align}\label{eq:features}
 \boldsymbol f=\big[&\ln(1+\max_k P_k),\ \ln(1+N^{-1}\sum_k P_k^2),\nonumber\\
 &\ln q_{.99}(P),\ \ln q_{.50}(P),\ \ln(1+N^{-1}\sum_n v_n^2),\nonumber\\
 &\max_r s_r-\min_r s_r,\ c_1,\ c_7,\ \iota,\ \sigma_{\mathrm{band}}\big]^{\mathsf T},
\end{align}
where $\iota=|\sum_n y_n^2|/\sum_n|y_n|^2$ measures complex impropriety and $\sigma_{\mathrm{band}}$ is the standard deviation across the four contiguous spectral-quarter means of $P_k$.

Auxiliary gate percentiles $s_{90}$ and $s_{10}$ are extracted from a unit-mean $65$-bin circular moving average of $|[\boldsymbol F\boldsymbol y]_k|^2$. Ordered gate execution evaluates binary predicates parameterized by the offline-frozen thresholds in Table~\ref{tab:rgparams}:
\begin{equation}\label{eq:gatepredicates}
\begin{gathered}
 q_0=\boldsymbol1\{d_c^2\leq\tau_d\},\\
 q_1=\boldsymbol1\{c_1<\tau_c,\ r>\tau_r,\ \iota<\tau_\iota\},\\
 q_2=\boldsymbol1\{c_1<\tau_c,\ p<\tau_{p,\mathrm{low}}\},\quad
 q_3=\boldsymbol1\{c_1<\tau_c\},\\
 q_4=\boldsymbol1\{p>\tau_{p,\mathrm{high}},\ s_{90}<\tau_{s90}\},\quad q_5=\boldsymbol1\{B=1\},\\
 q_6=\boldsymbol1\{r<\tau_r,\ B=2,\ a_4\geq\tau_{a4},\ \chi<\tau_\chi\},\\
 q_7=\boldsymbol1\{r<\tau_r,\ \chi\geq\tau_\chi,\ Q_2\lor Q_+\},\quad q_8=1,
\end{gathered}
\end{equation}
with $Q_2=\{B=2,a_4\geq\tau_{a4}\}$ and $Q_+=\{B>2,s_{10}\geq\tau_{s10}\}$, forming mutually exclusive indicators
\begin{equation}\label{eq:gatemasks}
 \pi_m=q_m\prod_{k<m}(1-q_k),\qquad\sum_{m=0}^8\pi_m=1,
\end{equation}
mapped sequentially to Core, local-power weighting, Identity, four-line projection, Core, unweighted $256$-point STFT, frequency Huber, unweighted $64$-point STFT, and Core. Under $1$-bit conversion where $y_n\in\{\pm1\pm\mathrm i\}$ has constant modulus, radial time Huber reduces to positive scalar multiplication $\mathcal H_\lambda(\boldsymbol y)=a\boldsymbol y$, leaving $q(\boldsymbol\theta,a\boldsymbol y)=a^2q(\boldsymbol\theta,\boldsymbol y)$ invariant and motivating transform-domain routing via $q_5$.

\begin{table}[!t]
\centering\footnotesize
\caption{Frozen Online Reliability Gate Parameters and Calibration Basis}
\label{tab:rgparams}
\setlength{\tabcolsep}{3.0pt}
\begin{tabularx}{\columnwidth}{@{}ll>{\looseness=0\raggedright\arraybackslash}X@{}}
\toprule
Symbol & Value & Offline calibration basis \\
\midrule
$\tau_d$ & $1.5$ & High-coverage quantile envelope of $d_c^2$ on upstream development set $\mathcal S_{\mathrm{rep}}$ \\
$\tau_c$ & $0.15$ & Bandlimited thermal and quantization lag-$1$ decorrelation ceiling \\
$\tau_r$ & $0.4$ & Quarter-block power-variation split between burst and stationary occupancy \\
$\tau_\iota$ & $0.1$ & Circularity ceiling isolating proper two-quadrature burst noise \\
$\alpha_p,\gamma_p$ & $0.2,1.8$ & Lower power quantile and Huber-equivalent attenuation scale in \eqref{eq:localpower} \\
$\tau_{p,\mathrm{low}}$ & $3.5$ & Periodogram peak split between weak broadband shifts and discrete tones \\
$\tau_{p,\mathrm{high}},\tau_{s90}$ & $7.0,0.8$ & Narrowband concentration envelope with uncorrupted spectral bulk \\
$\tau_{a4}$ & $0.8$ & $2$-bit log fourth-moment threshold above thermal quantizer occupancy \\
$\tau_\chi$ & $1/3$ & $64$-sample STFT concentration split between swept ridges and diffuse noise \\
$\tau_{s10}$ & $0.003$ & Lower smoothed spectral floor for conversion with $B>2$ \\
$\lambda_{\mathrm{RG}}$ & $1.345$ & Asymptotic $95\%$-efficiency Gaussian Huber constant for frequency clipping \\
\bottomrule
\end{tabularx}
\end{table}

\section{Non-Gaussian Quadratic-Score Covariance}\label{app:quadratic}
To capture quantization-induced non-Gaussianity in \eqref{eq:crossscore}, stack real and imaginary parts of $\boldsymbol z$ as $\boldsymbol r=[(\Re\boldsymbol z)^{\mathsf T},(\Im\boldsymbol z)^{\mathsf T}]^{\mathsf T}=\boldsymbol\mu+\boldsymbol u$ with zero-mean fluctuation covariance $\boldsymbol K_r=\E[\boldsymbol u\boldsymbol u^{\mathsf T}]$. Any Hermitian contrast-derivative matrix $\boldsymbol B_a$ maps to the real symmetric form
\begin{equation}
 \boldsymbol H_a=\begin{bmatrix}
 \Re\boldsymbol B_a&-\Im\boldsymbol B_a\\
 \Im\boldsymbol B_a&\Re\boldsymbol B_a
 \end{bmatrix},\qquad s_a=\boldsymbol r^{\mathsf T}\boldsymbol H_a\boldsymbol r.
\end{equation}
Denoting $\boldsymbol h_a=2\boldsymbol H_a\boldsymbol\mu$, third central moments $M_{ijk}=\E[u_i u_j u_k]$, and fourth joint cumulants $\kappa_{ijkl}$, centering and multiplying two quadratic scores yields
\begin{align}\label{eq:nongaussian}
 \Cov(s_a,s_b)&=\boldsymbol h_a^{\mathsf T}\boldsymbol K_r\boldsymbol h_b
 +2\tr(\boldsymbol H_a\boldsymbol K_r\boldsymbol H_b\boldsymbol K_r)\nonumber\\
 &+\sum_{ijk}(h_{a,i}H_{b,jk}+h_{b,i}H_{a,jk})M_{ijk}\nonumber\\
 &+\sum_{ijkl}H_{a,ij}H_{b,kl}\kappa_{ijkl}.
\end{align}
Stacking two distinct branch outputs into a joint real vector evaluates cross-branch score covariance identically, showing that empirical paired-error calibration absorbs skewness, kurtosis, and complex impropriety without estimating fourth-order sample tensors.

\bibliographystyle{ieeetr}

\end{document}